\documentclass{article}
\usepackage[utf8]{inputenc}
\usepackage{amsmath}
\usepackage{mathrsfs}
\usepackage{physics}
\usepackage{scalerel}
\usepackage{IEEEtrantools}
\usepackage{color}
\usepackage{soul}

\usepackage[title]{appendix}

\usepackage{subcaption}

\usepackage{graphicx}
\usepackage{sidecap}
\graphicspath{ {./images/} }
\usepackage[export]{adjustbox}

\newcommand{\pd}{\partial}

\newcommand{\bd}[1]{\mathbf{#1}}

\newcommand{\Eqn}[1]{Eq.~(\ref{#1})}

\usepackage{xcolor}

\DeclareRobustCommand{\hlp}[1]{{\sethlcolor{pink}\hl{#1}}}

\usepackage[numbers, sort&compress]{natbib}

\usepackage{algpseudocode}
\usepackage{algorithm}

\usepackage{authblk}
\author[1,2]{H. A. McDonough} 

\author[1,2]{Nicholas A. Mecholsky} 

\affil[1]{The Catholic University of America, 620 Michigan Ave
NE, Washington, 20064, DC, US}

\affil[2]{The Vitreous State Laboratory, Hannan Hall, CUA, Washington, 20064, DC, US}

\title{Modeling Energy- and Momentum-dependent Scattering Relaxation Times in a Semi-Classical Model of Charge Transport using the Self-Scattering Technique} 
\date{July 2025}

\begin{document}

\maketitle

\begin{abstract}
Improvement of numerical methods for calculating charge transport quantities of materials from the Boltzmann transport equation (BTE) is important for prediction of material properties. In particular, techniques which allow for more accurate models of scattering rates and band structures while remaining less computationally involved are valuable. The self-scattering technique is one such technique for implementing energy- and momentum-dependent scattering relaxation times in Monte Carlo simulations of charge transport or iterative techniques for solving the BTE. While the technique initially uses a constant relaxation time in the simulation algorithm, we found, upon analysis of the technique, that the energy and momentum total scattering relaxation time may be recovered. To show this, we preformed self-scattering Monte Carlo simulations of electron transport for both energy-dependent and independent relaxation times. The relaxation times and the fraction of scattering type selected in the simulation matched the full rates implemented in the simulation. In order to understand these results, we developed the theory behind the technique, demonstrating that the  probability distribution function of free-flight times created using the self-scattering algorithm produces the analytical probability distribution function expression generated by full energy, momentum and time dependent relaxation times and probabilities. However, certain forms of relaxation times may cause the simulation to fail, depending on implementation. Clarity about such techniques is important for improvement of charge transport simulation models and implementation of ab initio scattering rates and band structures.
\end{abstract}

\section{Introduction}
\label{sec:intro}

The Boltzmann transport equation (BTE) is of critical importance in studies of transport of various kinds, particularly electronic and thermal transport in semiconductor and thermoelectric materials. Its solution is the carrier distribution function in 6-dimensional real and momentum space, from which transport quantities such as the electronic conductivity may be calculated. The BTE may be given by \cite{Jacoboni1989}: 
\begin{equation}\label{eqn:BTE}
\frac{\pd{f}}{\pd{t}} + \boldsymbol{v}\cdot \grad_r f + \dot{\boldsymbol{k}} \cdot \grad_k f = \left.\frac{\pd f}{\pd t}\right|_{coll},
\end{equation}
where $f$ is the carrier distribution in real space and momentum, or $k$, space, and $\boldsymbol{v}$ and $\boldsymbol{k}$ are the carrier velocity and carrier wavevector. The term $\left.\frac{\pd f}{\pd t}\right|_{coll}$ corresponds to the contributions from collisions, i.e., from scattering events. The left-hand terms collectively are known as drift terms. The collision term for a transition from a carrier state represented by wavevector $\boldsymbol{k}$, to a state represented by $\boldsymbol{k}'$ may be given by \cite{Ashcroft_and_Mermin,Jacoboni1989}: 
\begin{equation}\label{eqn:fullcollision}
    \left.\frac{\pd f}{\pd t}\right|_{coll} = \frac{V}{(2\pi)^3} \int ( f(\boldsymbol{k}) R(\boldsymbol{k}, \boldsymbol{k}') (1 - f(\boldsymbol{k}')) - f(\boldsymbol{k}')R(\boldsymbol{k}',\boldsymbol{k})(1-f(\boldsymbol{k})) \dd \boldsymbol{k}',
\end{equation}
where $R(\boldsymbol{k}, \boldsymbol{k}')$ is the probability of transition per unit time, or transition probability rate, from a state $\boldsymbol{k}$, to a state $\boldsymbol{k}'$, and the terms $(1-f(\boldsymbol{k}))$ and $(1 - f(\boldsymbol{k'}))$ account for Pauli exclusion.

If this full, non-linear version of the BTE were solved analytically, it would provide a complete description of transport processes in a given system. However, the Boltzmann equation may only be solved analytically with the implementation of various approximations. One such approximation is the Relaxation Time Approximation (RTA). In this approximation, the full collision term integral is approximated by the difference between the full distribution, $f$, and the equilibrium distribution, $f_0$, divided by a \emph{scattering relaxation time}, $\tau$ \cite{Ashcroft_and_Mermin}:
\begin{equation}\label{eqn:RTA}
\left.\frac{\pd f}{\pd t}\right|_{coll} = \quad \frac{f - f_0}{\tau}.
\end{equation}
The relaxation time is sometimes taken to be a constant, as in the constant relaxation time approximation (CRTA), or may depend on the wavevector, $\boldsymbol{k}$, the energy of the pseudo-particle,  $\epsilon$, and through these, the simulation time, $t$.
The RTA, while widely implemented in Boltzmann solvers such as BoltzTraP and BoltzWann \cite{BoltzTraP2006,BoltzTraP22018,Boltzwann2014}, 
falls short when transport is very dependent on the carrier distribution function. The CRTA in particular has been shown to make incorrect predictions, and adds an arbitrary element into simulations, which would preferably be \textit{ab initio} \cite{jayaraj2022relaxation, Xu2014, Zhou2018, Sun2015}.

As an alternative to solving the full integro-differential BTE directly, numerical methods such as Monte Carlo simulations \cite{Jacoboni1983, Hammar1971, Jacoboni1989, Kunikiyo1994, Fischetti1988, Fischetti1991, Grondin1999, Bufler2000} and iterative \cite{Hammar1972, Hammar_1973, Nougier_Rolland_1973} techniques may be implemented. These methods solve the Boltzmann transport equation (BTE) numerically by assuming a starting distribution in position and momentum and allowing it to evolve over a series of time steps according to kinetic equations or path variables, as in Jacoboni et al. (1983) \cite{Jacoboni1983}.
In these techniques, the times selected for each evolution step are chosen from a scattering probability distribution in time, dependent on a \textit{total relaxation time}. This distribution is distinct from the position and momentum distribution which is the solution of the BTE. The total relaxation time is defined as the inverse of the sum of the scattering probability rates integrated over all possible final wavevectors of the carriers, as in Eqn.~\ref{eq:MR}. 
Various levels of complexity beyond the CRTA may be included in Monte Carlo simulations, depending on the accuracy of the dispersion or scattering rate models.

Realistic carrier relaxation times or scattering rates are dependent on the quasi-particle's path in $k$-space through the energy dispersion. Because of this, they have a complicated energy, momentum, time, and applied field dependence, which is determined by the band structure of the material in addition to other quantities relevant to the particular type of scattering (e.g.,~phonon wave-vectors and phonon band structures in scattering by lattice vibrations). The implementation of complicated expressions for the relaxation time within the collision term integral may be greatly simplified by implementing the \textit{self-scattering technique}.

The self-scattering technique was introduced by Rees in the 1960s \cite{rees1968, rees1969, rees1973} as a technique for use in iterative Boltzmann transport equation solvers, and has been widely implemented in Monte Carlo simulations of carrier transport since the 70s.
The self-scattering technique skips the evaluation of the total scattering integral at each time step by initially assuming a constant total scattering rate through the inclusion of a fictitious `self-scattering' scattering mechanism in addition to energy-dependent scattering rates. Using a constant relaxation time in the Monte Carlo framework causes the distribution from which the free-flight times are selected to simplify to an exponential distribution, greatly simplifying the procedure. After performing the simulation, the free-flight times are retroactively assigned scattering types and new carrier states based on the full collision expression and relaxation time expressions. The initial process randomly selects free-flight times from an exponential distribution, but after post-processing, as we show in section \ref{sec:dists}, recreates the true energy, momentum, and time dependent scattering probability distribution of the total relaxation time.

Justification for the self-scattering technique is sadly lacking in the literature. Given the current interest in more accurate models of scattering relaxation times and prediction of charge transport quantities such as the conductivity and carrier mobility from ab initio calculations \cite{jayaraj2022relaxation, Ganose2021, guistino2017, Li2021, Graziosi2020},
this technique merits re-examination and explication. The simplicity of the technique makes it an attractive option for use in Monte Carlo calculations of charge transport quantities from density functional theory (DFT) as in \cite{Li2013}, or from other \emph{ab initio} band structures and relaxation times.  

In this paper, we re-examine the workings of the self-scattering technique within the context of the \emph{semi-classical} Monte Carlo framework from Jacoboni et al. (1983) \cite{Jacoboni1983}, and show that it re-produces accurate scattering relaxation times and probability distributions for the scattering mechanisms simulated. 
We also mention caveats that must be taken into account when using the self-scattering technique.
In section \ref{sec:semiclass}, we explain the basics of the Monte Carlo simulation model; in \ref{sec:selfscatter}, we explain the implementation of the self-scattering technique; in section \ref{sec:expverf}, experimental simulation results are examined and the technique is shown to recover the total analytical relaxation time from Matthiessen's rule; in section \ref{sec:dists}, the theory behind the free-flight time distributions created by self-scattering is explained; and these concepts and some limitations are finally discussed in section \ref{sec:discuss}.

\section{The Monte Carlo Transport Model}\label{sec:semiclass}
The Monte Carlo transport simulation uses a \textit{semi-classical} transport model. 
In this section we review general features of our simulation and scattering models, but for more details on Monte Carlo simulation within the semi-classical Boltzmann transport framework, please refer to Jacoboni et al. \cite{Jacoboni1983, Jacoboni1989}.
Semi-classical carrier transport models the carrier as a Bloch wave function in a periodic crystal lattice, given by:
\begin{equation}
    \psi_k(x)=u_k(x) e^{i k x}. 
\end{equation}
In this one-dimensional expression, $k$ is the carrier wavenumber and $u_k$ is a slowly varying function that has the same periodicity as the crystal lattice. This form identifies the carrier as a modified plane wave of a single wavenumber $k$. 
The periodicity of crystalline solid potentials allows the carrier to be modeled as a free quasi-particle because the dispersion is only dependent on the wavenumber $k$. 

A simplified semi-classical model incorporates quantum mechanical scattering processes but isolates the quantum mechanical effects and leaves the intra-quantum mechanical propagation to classical mechanics. This model of the carrier is implemented in Monte Carlo simulations of transport as a series of classical free-flights interspersed by quantum mechanical scattering events, where the carrier interacts with a perturbing scattering potential. During free-flight, an electron under the influence of an applied electric field, $\boldsymbol{E}$, changes its momentum according to the classical expression: 
\begin{equation}\label{eqn:kdot}
    \hbar \Dot{\boldsymbol k} = -|e| \boldsymbol E,
\end{equation}
from the de Broglie relation and Newton's law. 
The duration of each free-flight may then be determined either by selecting time variables from a probability distribution derived from the total scattering probability rate, or by using the self-scattering technique.

Scattering probability rates for transitions between electron states $\boldsymbol k$ and $\boldsymbol{k}'$ and crystal states $c$ and $c'$ may be calculated from Fermi's Golden Rule:
\begin{equation}\label{FGR}
    R(\boldsymbol k, c ;\boldsymbol{k}', c') = \frac{2\pi}{\hbar}\left|\bra{\boldsymbol{k}', c'}
    H'\ket{\boldsymbol{k}, c}\right|^2 \delta [\epsilon (\boldsymbol{k}', c') -\epsilon (\boldsymbol{k},c)],
\end{equation}
where $H'$ is the perturbation Hamiltonian for the type of scattering interaction being considered, $\delta$ is the Dirac delta function, and $\epsilon(\boldsymbol k, c)$ represents the energy dispersion of the carrier for that state. These scattering rates may be dependent on many factors, because of the path dependence of the carrier's energy through the Brillouin Zone (BZ). These variables include the carrier energy (through the dispersion relation), wavevector, time, and applied fields. 
In this paper, we concentrate on the energy dependence of the scattering rates in the our experimental exploration, Section~\ref{sec:expverf}, and wave-vector and time dependence in the analytical examination of the technique, Section~\ref{sec:dists}. 

The calculation of the matrix element in \Eqn{FGR} involves some difficulty. It may be taken from empirical fits as in early Monte Carlo simulations \cite{Jacoboni1983}, or, as in more recent approaches, calculated through \textit{ab initio} methods such as in sources \cite{Li2021, guistino2017} for the electron-phonon matrix element. 

To determine a free-flight time at a given simulation step, i.e., between scattering events, any number of scattering probability rates may be combined using Matthiessen's Rule to form the total scattering rate $R$. The total relaxation time, $\tau$, is defined as the inverse of $R$:
\begin{equation}\label{eq:MR}
     R = \frac{1}{\tau} = \frac{1}{\tau_1}+\frac{1}{\tau_2}+\frac{1}{\tau_3} \cdots,
\end{equation}
where each $\tau_i$ is a scattering relaxation time, the inverse of a given scattering probability rate integrated over final states $\boldsymbol{k}'$, and is dependent on the same parameters, e.g.,\ the carrier wavevector and simulation time:
\begin{equation}
    \frac{1}{\tau_1} = R_1[\boldsymbol k(t)] = 
    \int R_1[\boldsymbol{k}'(t), \boldsymbol{k}(t)] \dd \boldsymbol{k}'.
\end{equation}
That scattering probability rates may be added in this way may be seen intuitively when thinking about rates of collision\footnote{As an example, if it is both raining and lightly hailing, the rate of feeling either a raindrop or a hailstone on your outstretched finger would be the sum of both the rate of a hailstone or of a raindrop since it is very unlikely to have both at the same time.}.

The total scattering rate may then be used to calculate the probability of a carrier scattering within $\dd t$ of time $t$ when traveling classically from an initial wavevector:
\begin{equation}\label{AnalyticalExpression}
    \mathscr P(t) dt =  R[\boldsymbol k(t)] \exp \left( -\int_0^t R[\boldsymbol k(t')] dt' \right) dt,
\end{equation}
where $R[\boldsymbol k(t)] = \sum_j R_j[\boldsymbol k(t)]$ is the total scattering rate, the sum of all the integrated scattering probability rates included in the simulation.
A free-flight time, that is, the length of time the carrier is supposed to travel classically before scattering at each simulation step, may then be selected randomly from Eqn.~\ref{AnalyticalExpression}. This scattering probability, $\mathscr{P}$, 
will depend on the carrier energy, wavevector and time through the scattering rates. Because of this, we sometimes refer to $\mathscr{P}$ in relation to various functions or variables that it depends on, as in $\mathscr{P}(\epsilon)$, or $\mathscr{P}[\boldsymbol{k}(t)]$.

After a free-flight in the simulation, a scattering event occurs, and the type of scattering which took place is selected randomly, weighted by the relative ratios of scattering rates for that point in $\bd{k}$-space. After scattering, the new state in $\bd{k}$-space is chosen according to a distribution derived from the selected scattering mechanism, and a new free-flight begins. At each scattering event, information about the carrier's state: the wave vector, energy, velocity, scattering mechanism type and free-flight time, is collected, and the simulation continues until sufficient data has been gathered.

The main computational difficulty in this simulation scheme arises from the calculation of the total scattering probability integral, Eqn.~\ref{AnalyticalExpression}, at various values of $\boldsymbol{k}$ in the simulation. Self-scattering allows this process to be greatly simplified by setting the total rate to a constant for the purposes of free-flight time selection, which changes the probability expression into a simple exponential distribution in time. The selection of scattering type for a given free-flight may also be partially simplified, by only requiring certain scattering rates up to the one chosen to be evaluated. Self-scattering is explained more fully in the following section.

\section{The Self-Scattering Technique}\label{sec:selfscatter}

In the self-scattering technique, the wavevector, time, and other variable dependent sum of scattering rates in \Eqn{AnalyticalExpression} is initially replaced by a constant total, $\Gamma$ or $\frac{1}{\tau_0}$. This estimated distribution is later replaced by an energy, time, and/or 
wavevector-dependent probability distribution by a post-processing step. 

With the substitution of a constant relaxation time, \Eqn{AnalyticalExpression} becomes an exponential distribution:
\begin{equation}\label{GammaExponentialEquation}
        P_\Gamma(t) = \Gamma e^{-\Gamma t}.
\end{equation}
The electron free-flight time may be chosen from an exponential distribution of times according to the equation: 
\begin{equation}
    t_r = -\tau_0 \ln(r),
\end{equation}
where $r$ is a uniformly chosen random number between 0 and 1.
After the free-flight duration is selected, the type of scattering which occurred at the end of the free-flight is selected in a manner that respects the relative frequencies of scattering.

To make this selection, a uniformly chosen random number, $y$, between 0 and $\Gamma$, is chosen, and sequential sums of the scattering rates are evaluated at the carrier's energy at the given scattering event. The random number $y$ is then compared to each sum in turn, and
the scattering mechanism which was added to make the total sum of the rates greater than $y$ is chosen as the type of scattering which took place. This allows for evaluation of only the scattering probability rates included in the sum up until the scattering mechanism chosen, and avoids the final integral over the total scattering probability rate.

\begin{figure}
    \centering
    \includegraphics[width=.85\textwidth]{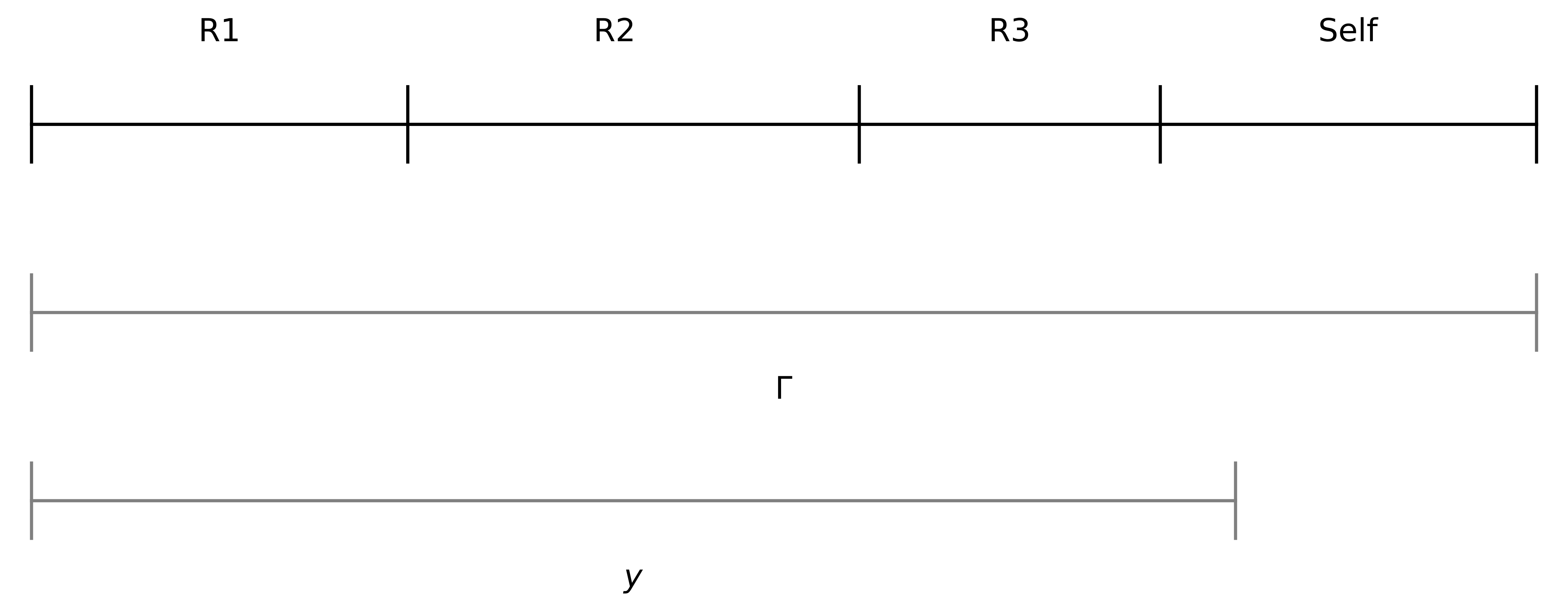}
    \caption{Technique for choosing among scattering mechanisms. The total scattering probability rate is set equal to a constant, $\Gamma$, represented by the second line in the figure and should be chosen to be larger than the sum of all scattering rates in the region of interest. The contributions to the total scattering rate (Rate 1, Rate 2, Rate 3, and Self-scattering) are represented by the sections of the top line ($R_1$, $R_2$, $R_3$, and $Sel \! f$), with Self-scattering making up any difference between the total constant rate and the sum of the variable-dependent rates.
    If $\sum_{i=1}^{j-1} R_i < y < \sum_{i=1}^{j} R_i$, then $R_j$ is selected as the scattering rate. If $\sum_{i=1} R_i < y \le \Gamma$, then self-scattering is chosen.  In this figure, self-scattering will be selected.
    }
    \label{fig:mechselect}
\end{figure}

If $y$ is greater than the sum of all of the scattering rates, then “self-scattering” is said to have taken place, which is a fictitious scattering event representing a continuation of the free-flight rather than a true scattering event. This selection method is illustrated in Fig.~\ref{fig:mechselect}. The top line of the figure indicates the contributions of scattering rates, $R_1$, $R_2$, and $R_3$, and the additional contribution of self-scattering to the total constant scattering rate, $\Gamma$, indicated by the second line. The third line is an example quantity for $y$, which falls between the sum through $R_3$, and the total $\Gamma$, thus choosing self-scattering rather than one of the scattering rates.

After the selection of the scattering type, the free-flight times before the fictitious self-scattering events may be added to the antecedent true events' free-flight times. In this way, an accurate free-flight time distribution may be reconstructed from the data. 

It is not immediately clear that this addition of self-scattering as a fictitious scattering rate will leave the ratio between the numbers of different types of scattering events chosen, the relaxation times of the scattering mechanisms, or the total scattering relaxation time unchanged. The rest of the paper is dedicated to the exploration of the influence of the self-scattering method on reproducing the effect of various scattering mechanisms. In sections \ref{sec:expverf} and \ref{sec:dists}, the distribution of free-flight times created by this method is shown to coincide with the distribution created by the total relaxation time rather than individual relaxation times of scattering mechanisms, and the ratios of scattering probabilities as a function of energy are shown to be reflected by how frequently each type of scattering is chosen within the simulation. When this method is implemented, the total scattering rate, $R$, or total relaxation time, $\tau$, may be recovered from simulated data, but not the individual relaxation times of particular scattering mechanisms.

\section{Experimental Demonstration}
{\label{sec:expverf}}

To explore the self-scattering mechanism in more detail, we will start with a model that is complicated enough to capture the features of multiple scattering processes. We performed simulations of electron transport in $n$-doped silicon using the semiclassical Monte Carlo approach to solving the Boltzmann equation. We employed the self-scattering technique to investigate its effect on the total probability distribution from \Eqn{AnalyticalExpression}, and to confirm that each scattering mechanism is selected by self-scattering with the correct relative frequency.  
Pseudocode for the simulation is given below in Algorithm~\ref{alg:Sim}, with variable assignments for the pseudocode given in Table~\ref{tab:pseudocode_variables}.
\begin{algorithm}[hbt!]
\caption{Self-Scattering Monte Carlo Simulation}\label{alg:Sim}
\begin{algorithmic}

\State $\bd{begin}$

\State $j \gets 1$

\State $record = ()$
\While{$j < trajnum$}
\State reset initial conditions: 
\State\hspace{\algorithmicindent} select $\epsilon_0$ from $f_D(T)$ 

\State\hspace{\algorithmicindent} $\boldsymbol{k}_0 \gets \boldsymbol{k}(\epsilon_0) $

\State\hspace{\algorithmicindent} $n \gets 1 $

\While{$n > trajlen$}
\State $\tau_{\Gamma} \gets 1/\Gamma$

\State $t_{f\!f} \gets - \tau_{\Gamma} \ln{rand_1}$

\State $\boldsymbol{k}_1 \gets \boldsymbol{k}_0 -\frac{|e|}{\hbar} \boldsymbol{E} \ t_{f\!f}$

\State $\epsilon_1 \gets \epsilon(\boldsymbol{k}_1)$

\State find successive sums of scattering rates: 

\State\hspace{\algorithmicindent} $S_1 \gets R_1(\epsilon_1)$, $S_2 \gets R_1(\epsilon_1) + R_2(\epsilon_1)$, $\cdots$ etc.

\If{$S_{tot} = \sum_i R_i(\epsilon_1) > \Gamma$},
$\Gamma \gets S_{tot}$,

\EndIf

\State $y \gets rand_2 \ \Gamma$

\State $R_{f\!f} \gets $1st $R^*$ with $ S^* > y$,

\State $\boldsymbol{k}_3 \gets \boldsymbol{k}(R_{ff})$

\State $\boldsymbol{k}_0 \gets \boldsymbol{k}_3$

\State $record \ +\!= (\boldsymbol{k}_3$, $\epsilon_1$, $R_{f\!f}$, $t_{ff})$
\State $n \ +\!= 1$
\EndWhile
\State $j \ +\!= 1$
\EndWhile
    
\end{algorithmic}    
\end{algorithm}
\begin{table}[t]
    \begin{subtable}[t]{0.49\textwidth}
    \flushleft
    \begin{tabular}[t]{| p{0.23\textwidth} | p{0.63\textwidth} |}
            \hline
            \centering Variable & Assignment \\[3pt] \hline \hline
            \centering $trajnum$ & The desired number of carrier trajectories \\[3pt]
            
            \centering $\epsilon_i$ & A carrier energy \\[3pt]
            
            \centering $f_D(T)$  & The Fermi-Dirac distribution as a function of temperature, $T$  \\[3pt] 
            
            \centering $\boldsymbol{k}_i$ & A carrier wavevector \\[3pt]
            
            \centering $\boldsymbol{k}(\epsilon)$ & The carrier wavevector as a function of energy, $\epsilon$ \\[3pt] 
            
            \centering $trajlen$ & The desired number of free-flights or scattering events in a trajectory \\[3pt]
            
            \centering $\tau_{\Gamma}$ & The constant, self-scattering, total relaxation time \\[3pt]
            
            \centering $\Gamma$ & The exponential parameter of the self-scattering free-flight time distribution \\[3pt]
            
            \centering $t_{f\!f}$ & The free-flight time \\[3pt]
            
            \centering $rand_i$ & A random number between 0 and 1 \\[3pt]
            
            \centering $e$ & The elementary charge \\[13pt]
            \hline
        \end{tabular}
    \end{subtable}
    \hspace{\fill}
    \begin{subtable}[t]{0.49\textwidth}
    \flushright
    \begin{tabular}[t]{| p{0.23\textwidth} | p{0.63\textwidth} |}
        \hline
        \centering Variable & Assignment\\[3pt] \hline \hline
        
        \centering $\boldsymbol{E}$ & The applied electric field vector \\[3pt]
        
        \centering $\hbar$ & The reduced Planck's constant \\[3pt]
        
        \centering $\epsilon(\boldsymbol{k})$ & The energy dispersion function \\[3pt]
        
        \centering $S_N$ & $\sum_{i = 0}^N R_i(\epsilon)$ \\[3pt] 
        
        \centering $R_i(\epsilon)$ & The integrated scattering rate of scattering mechanism $i$ as a function of carrier energy \\[3pt]
        
        \centering $S_{tot}$ & $\sum_i R_i(\epsilon)$ \\[3pt]
        
        \centering $y$ & A random fraction of $\Gamma$ \\[3pt]
        
        \centering $R^*$ & The critical scattering rate whose corresponding sum is greater than $y$ \\[3pt]
        
        \centering $R_{f\!f}$ & The scattering rate representing the true scattering type chosen to end the free-flight \\[3pt]
        
        \centering $\boldsymbol{k}(R)$ & The carrier wavevector as a function of the scattering mechanism represented by $R$ \\[3pt]
        \hline
    \end{tabular}
    \end{subtable}
    \caption{Variables used in Algorithm~\ref{alg:Sim}}
    \label{tab:pseudocode_variables}
\end{table}

In order to capture some of the complexity of multiple, realistic scattering mechanisms, in addition to simulations with constant relaxation times, we included energy dependent scattering rates for ionized impurity scattering and inelastic acoustic phonon scattering from Jacoboni et al. (1983) \cite{Jacoboni1983}. Expressions for these scattering rates are given in \ref{appendixA}. The values for the constant relaxation time were taken from the energy dependent rates for elastic phonon scattering and ionized impurity scattering with the energy, $\epsilon$, set equal to 1 eV.

For both constant and energy dependent relaxation times, material parameters were taken for silicon with a single parabolic band minimum from Ref.~\cite{Jacoboni1983}.  
The experimental parameters used were a temperature of $300 \ \mbox{K}$, an applied electric field of $100 \ \mbox{V}/\mbox{cm}$ in the $x$-direction in real space, and an arsenic dopant concentration of $10^{15} \ \mbox{cm}^{-3}$. These parameters were taken as typical room-temperature conditions for Si in a region where both scattering mechanisms are active.
The initial energy and momentum states of the electron for the energy-dependent case were randomly selected from the Fermi Dirac distribution at 300 K. 
For the energy-dependent and energy-independent cases, 750 trajectories of 5000 and 2000 free-flights respectively, were generated, and the free-flight times associated with a particular type of scattering were collected.

A post-processing step was included where the free-flight times before a self-scattering event were added onto the free-flight times of the next true scattering event in order to eliminate the effects of the fictitious self-scattering event.  In this way, self-scattering events were distributed to the other two legitimate scattering types. 

The data were sorted first by scattering type, then into pairs of free-flight times and energies. These data pairs were then sorted into energy bins of width 0.0005 eV between 0.0005 eV and 0.3005 eV.
\begin{figure}
    \centering
    \includegraphics[width=0.95\linewidth]{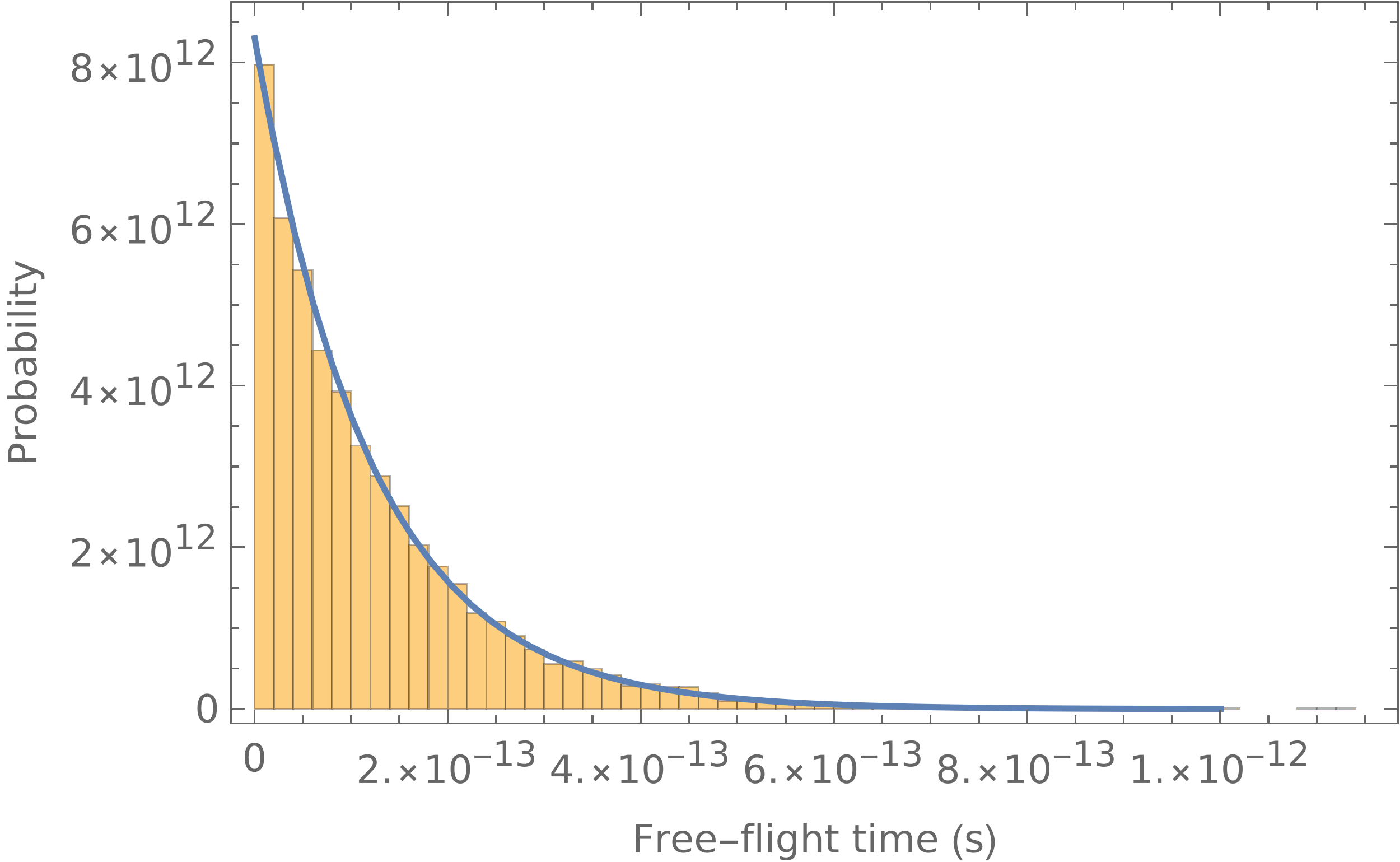}
    \caption{PDF Histogram with exponential fit,  $P(t)= (1/\tau) \exp(- (1/\tau) t)$, where $\tau = 1.20*10^{-13} \ $s, shown in blue for the free-flight times of 8824 phonon absorption scattering events occurring with energies between 0.002 eV and 0.0025 eV. \hlp{}
}
    \label{fig:expfit4}
\end{figure}
Within these small energy windows, the energy-dependent relaxation time will be approximately constant, and therefore a histogram of free-flight times chosen at this energy will appear exponentially distributed. Within each of these bins, the distribution of the free-flight times was examined, fit to an exponential distribution, and the relaxation time was taken to be the exponential parameter for that bin. An example PDF histogram of the pre-scattering free-flight times of 8824 energy-dependent phonon absorption scattering events within a bin of range, 0.002 eV to 0.0025 eV is shown in Fig.~\ref{fig:expfit4}. The exponential fit with resulting scattering relaxation time $\tau =  1.20*10^{-13} \ \mbox{s}$ is plotted on top in blue.

\begin{figure}
    \centering
    \includegraphics[width=\linewidth]{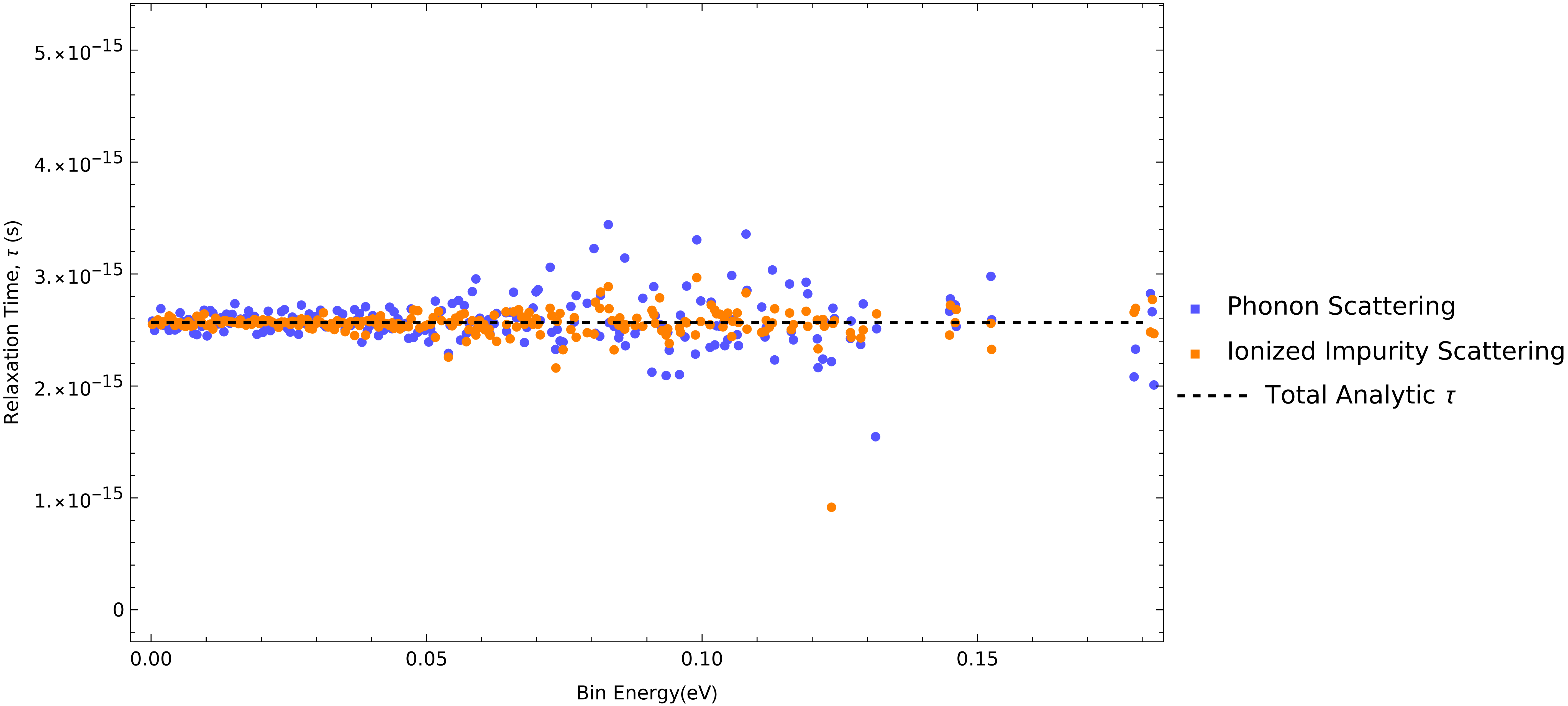}
    \caption{Relaxation times calculated from exponential fits of free flight time distributions corresponding to events occurring within energy bins of width 0.001 eV. Relaxation times calculated for constant scattering using parameters from phonon scattering are shown in blue, while values for constant scattering using parameters from ionized impurity scattering are shown in orange. The total analytic relaxation time, $\tau = 2.57\times 10^{-15}$, is shown as a black dashed line. A greater departure from the analytic relaxation time is expected at higher energies because of the reduced likelihood of higher energy states occurring within the simulation.
    }
    \label{fig:const_tau}
\end{figure}

\begin{figure}
    \centering
    \includegraphics[width = \linewidth]{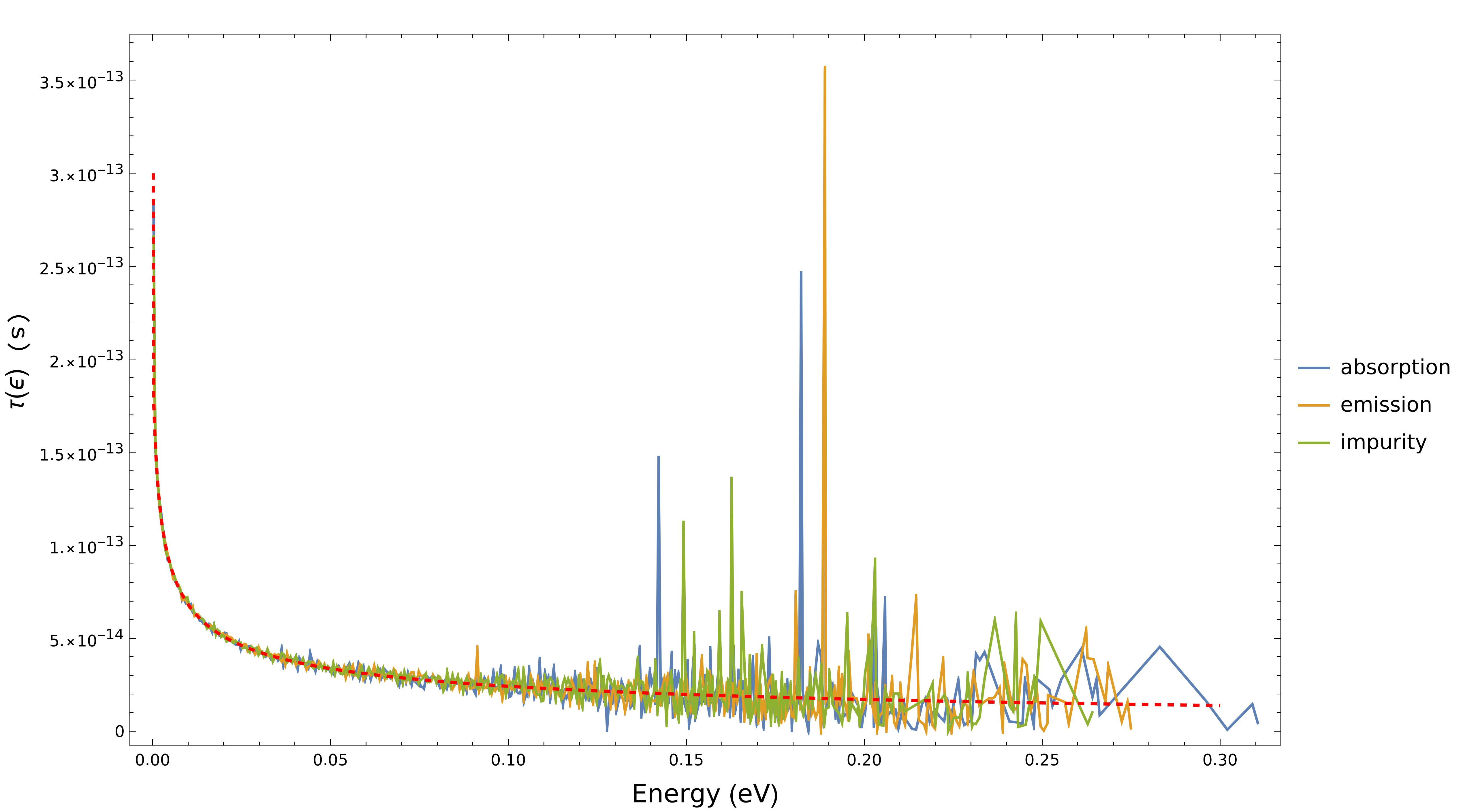}
    \caption{Simulated energy dependent relaxation times for inelastic acoustic phonon scattering and ionized impurity scattering as a function of average energy, and the total energy dependent analytic relaxation time. The relaxation times were calculated by fitting the distribution of free-flight times within a bin of width 0.0005 eV to an exponential distribution and taking the exponential parameter as the relaxation time for that bin. The relaxation times for inelastic absorption and emission mechanisms are plotted in blue and orange respectively, and impurity scattering is plotted in green. The data points for each relaxation time type are connected as a continuous line for better visibility. The total \emph{a priori} energy-dependent analytic relaxation time is shown as a red dashed line. Note that the higher disagreement at higher energy values arise from fewer data values associated with those unlikely energy states.} 
    \label{fig:tauofe}
\end{figure}

Each of the exponentially fit relaxation times was plotted against the average bin energy for each type of scattering and compared to the total relaxation time from \Eqn{eq:MR} as a function of energy for the scattering types in the simulation. This is shown in Fig.~\ref{fig:const_tau} for the energy-independent case and Fig.~\ref{fig:tauofe} for the energy-dependent case. 
In Fig.~\ref{fig:const_tau}, values for relaxation times calculated from exponential fits of the scattering probability distribution within narrow energy bins are plotted as blue dots for a constant elastic phonon scattering rate and as orange dots for a constant ionized impurity rate. The constant sum of these two rates is shown as a black dotted line on top. The relaxation time recovered appears to be within $5\times 10^{-16}$ s of the correct rate for values lower than 0.10 eV, with the value improving at lower energies.
In Fig.~\ref{fig:tauofe}, the values calculated for scattering by phonon absorption, phonon emission, ionized impurities are shown by blue, orange and green lines, respectively, with the total input relaxation time is plotted as a function of energy as a dashed red line on top. 
Note that the analytical total scattering relaxation time fits any of these relaxation time plots very well, while relaxation times corresponding to the individual rates are not recovered by the simulation. The disagreement towards the higher energy end of the plot comes from a lack of data, because it is less likely for the carrier to reach these higher energy states.

\begin{figure}
    \includegraphics[width= \textwidth]{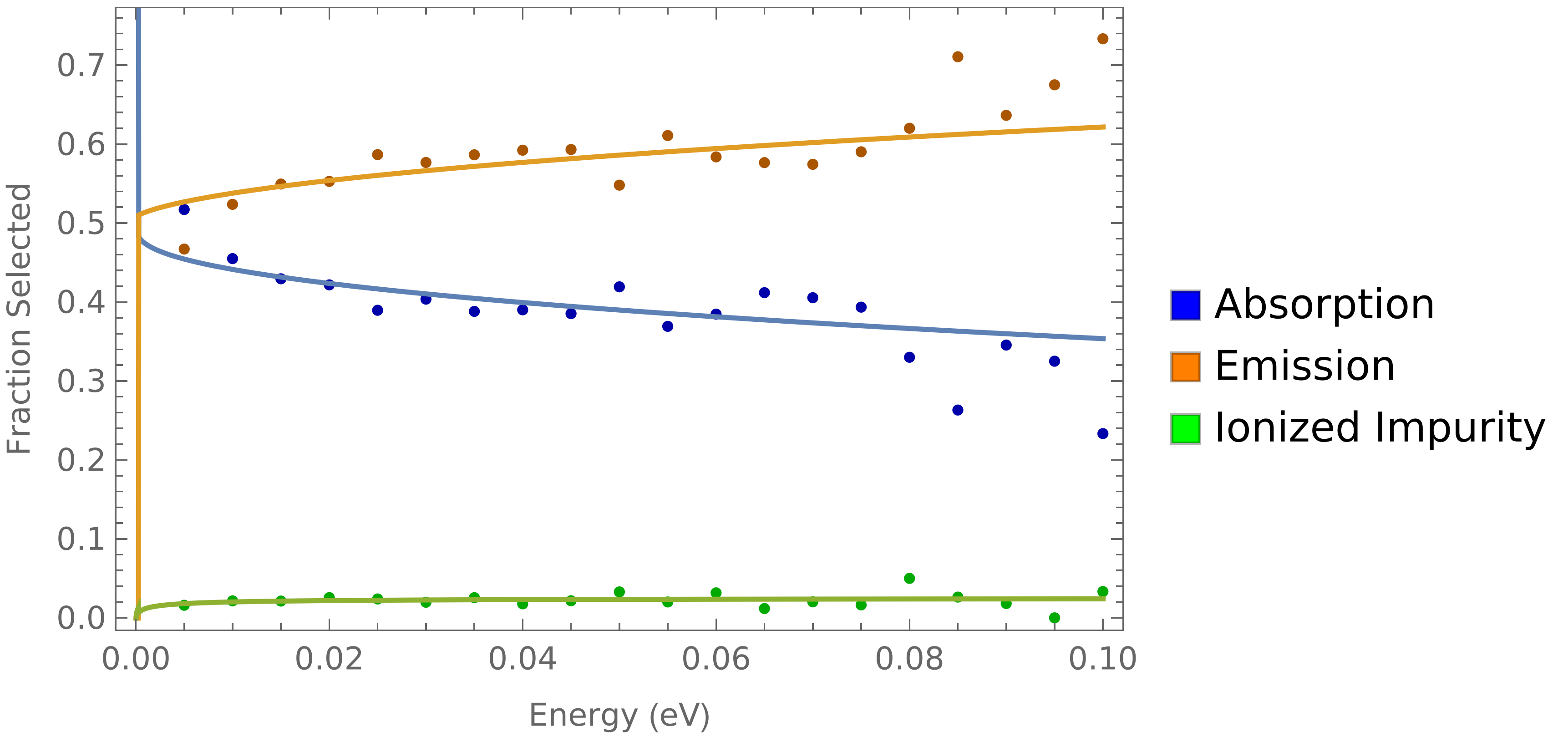}
    \caption{Fractions of scattering types selected as a function of energy. Individual scattering rates were divided by the total scattering rate and these expressions were plotted as a function of energy. These analytic fractions of phonon absorption, phonon emission, and ionized impurity scattering were plotted as blue, orange, and green curves, respectively.
    The expressions used for the scattering rates are given in \ref{appendixA}.
    Dots of the same colors represent the fraction of simulated scattering events of the same type within a given energy bin of size 0.005 eV. The parameters used for the simulation and rates were a temperature of 300 K, an applied field of 10,000 V/m, an ion concentration of $10^{15}\  \mbox{cm}^{-3}$ and 100,000 free-flights.}
    \label{fig:ratiofig}
\end{figure}

In order to confirm that particular types of scattering were selected with the correct frequency within the energy-dependent model, we also took ratios of the number of simulated scattering events of a particular type to the total number of simulated scattering events occurring within an energy bin. These were plotted against the average energy in each bin and compared with the ratios of the analytical scattering rates to the total scattering rate. This is shown in Fig.~\ref{fig:ratiofig}. 
In that figure, continuous curves show each scattering rate function divided by their total sum. Simulated phonon absorption scattering event ratios are plotted as dark blue dots on top of a blue line, phonon emission event ratios are plotted as dark orange dots on top of an orange line and ionized impurity event ratios are plotted as green dots on top of a light green curve. The data appears to closely approximate the analytical scattering rate expressions, with the greater variance at higher energies arising from the smaller number of scattering events in the higher energy bins.

These results indicate that in an energy dependent, self-scattering simulation, the relaxation times follow the probability function generated by Matthiessen's Rule from energy dependent scattering rates, Eqn.~\ref{AnalyticalExpression}, and that the the different types of scattering were selected with the correct relative frequency. Thus, in these simulations self-scattering appears to be a valid technique.

\section{The Self-Scattering Probability Distribution}{\label{sec:dists}}

Having seen the experimental validity of the self-scattering technique, let us examine the self-scattering process from an analytical perspective to determine the final distribution of free-flight times. For simplicity, we will begin with the assumption of single scattering type with a probability rate which is constant in time. 
Suppose $p$ is the probability of a ``true" scattering event (a non-self-scattering event) occurring at the end of a free-flight. We will call the ends of these free-flights, ``collision events" or ``collisions" in the following, although they do not necessarily involve a true scattering event. It follows that $(1-p)$ will be the probability that a true scattering event does not occur at a collision, meaning that self-scattering occurs. 

Within the semi-classical, self-scattering framework, the electron's trajectory continues until a true scattering event occurs, at which point the electron begins a trajectory in a new direction with a new velocity. 
There may be any number of self-scattering events in a trajectory, but only one true scattering event at the end of the trajectory. 
Writing out all of the possibilities, the probability that a true scattering event occurs at the first collision will be $p$, the probability that it occurs at the second collision will be $(1-p)p$, the probability that it occurs at the third collision will be $(1-p)^2 p$ and so on until $(1-p)^n p$, where $n$ is the number of free-flights, or the number of collision events in the trajectory. Thus, the probability of a successful scattering event will be distributed geometrically.
Summing this series to infinity gives a total probability of 1, as expected, since the true scattering event will occur at some point:
\begin{equation}
    \sum_n^\infty p (1-p)^n = 1.
\end{equation}

Now consider an additional independent scattering mechanism so that we have two scattering types: type 1 and type 2. Let these scattering event types have constant scattering probability rates $R_1$ and $R_2$ respectively. From the previous example, we know that the probabilities for a true collision event are distributed geometrically. In the self-scattering technique, we take our independent free-flight times from an exponential distribution with parameter $\Gamma > R_1 + R_2$. With this consideration, $\frac{\Gamma -R_1 -R_2}{\Gamma}$ will be the probability of a self-scattering event occurring, and $\frac{R_1}{\Gamma}$ will be the probability of scattering type 1 occurring, for instance. 
Focusing momentarily on scattering type 1, the probabilities of 0, 1, 2, or $n$ self-scattering events occurring, and then type 1 scattering occurring will be $\frac{R1}{\Gamma}$, $\frac{\Gamma -R1 - R2}{\Gamma}(\frac{R1}{\Gamma})$, $ (\frac{\Gamma - R_1 - R_2}{\Gamma})^2(\frac{R1}{\Gamma})$, and $(\frac{\Gamma -R_1 - R_2}{\Gamma})^n(\frac{R1}{\Gamma})$, respectively.

\begin{figure}
    \centering
    \includegraphics[width = \linewidth]{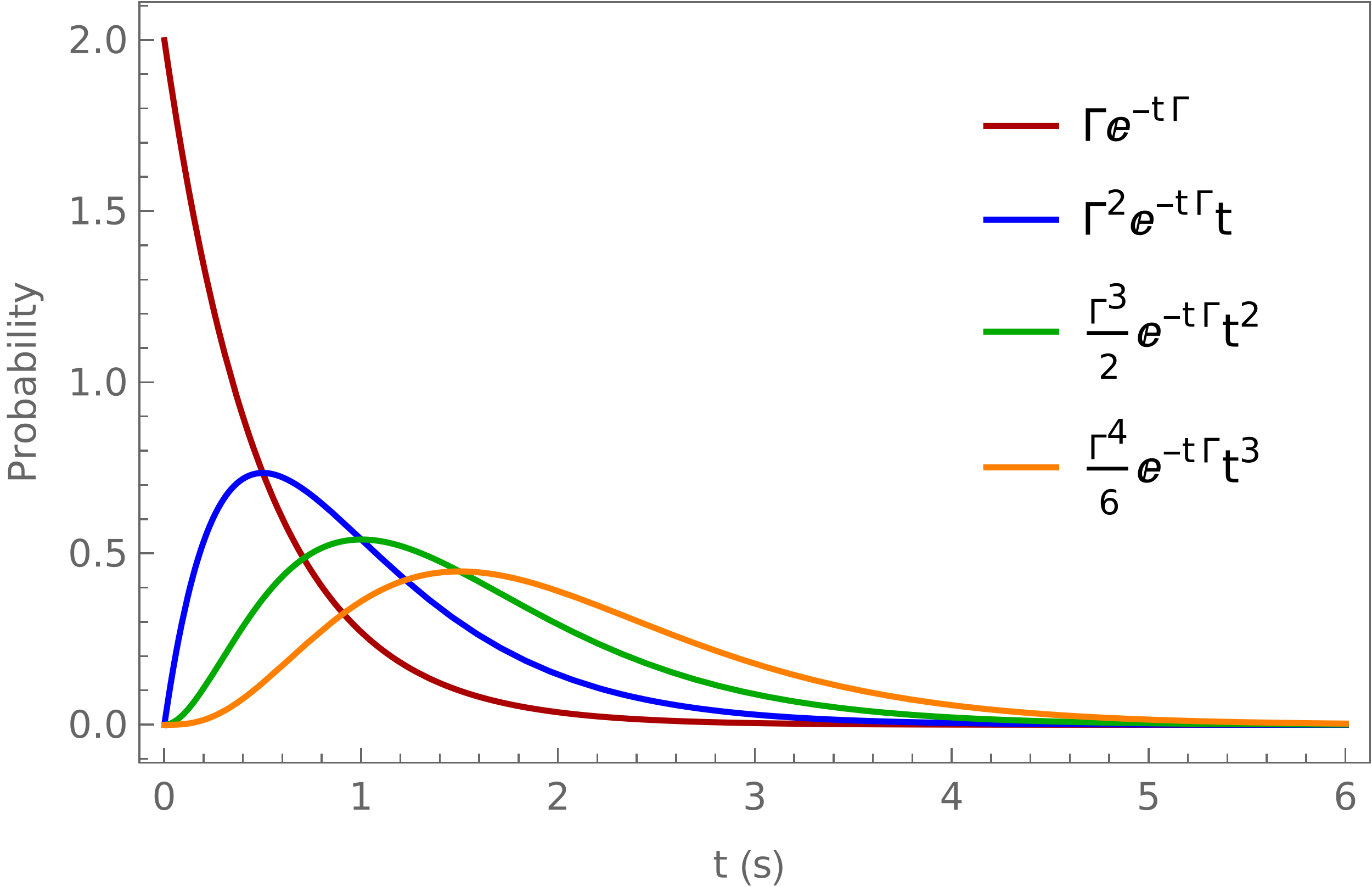}
    \caption{The first four Erlang distributions, with $\Gamma = 2$, shown in red, blue, green, and orange respectively. These functions weighted by the probabilities of real and fictitious self-scattering events combine to form a mixture distribution which gives the final probability distribution in self-scattering, as shown in Fig.~\ref{fig:ConstRateProbSum}
    }
    \label{fig:erlangs}
\end{figure}

Within the self-scattering framework, the individual free-flight times preceding each true or fictitious scattering event are chosen from the assumed exponential distribution, \Eqn{GammaExponentialEquation}. Thus, after post-processing, some true events will have free-flight times chosen directly from this exponential distribution (when there is no self scattering), while some will be a sum of two or more random variables chosen from this parent exponential distribution. The distribution which corresponds to a set of variables randomly generated from a sum of exponentially distributed variables is the Erlang distribution. The general formula for the Erlang distribution, using the variables for our case, is: 
\begin{equation}
    d_k(t) = \frac{\Gamma^{k+1} t^{k} e^{-\Gamma t}}{k!}.
\end{equation} 
Where $d_k$ is the probability distribution function (PDF) of times $t$ where $\Gamma$ is a rate parameter and $k+1$ is the number of exponential random variables added together to create each total random variable in the distribution. For our scenario, $k$ is the number of self-scattering events before scattering. 
The PDFs of the first four Erlang distributions are given by:
\begin{equation}
    \begin{array}{cc}
     d_0(t) = e^{- t \Gamma} \Gamma,    &  d_1(t) = e^{-t \Gamma} t \Gamma^2, \\[4pt]
    d_2(t) = \frac{1}{2} e^{-t \Gamma} t^2 \Gamma^3,     & d_3(t) =\frac{1}{6} e^{-t \Gamma} t^3 \Gamma^4.
    \end{array}
\end{equation}
Plots of these distributions are given in Fig.~\ref{fig:erlangs}.
\begin{figure}
    \centering
    \includegraphics[width=\linewidth]{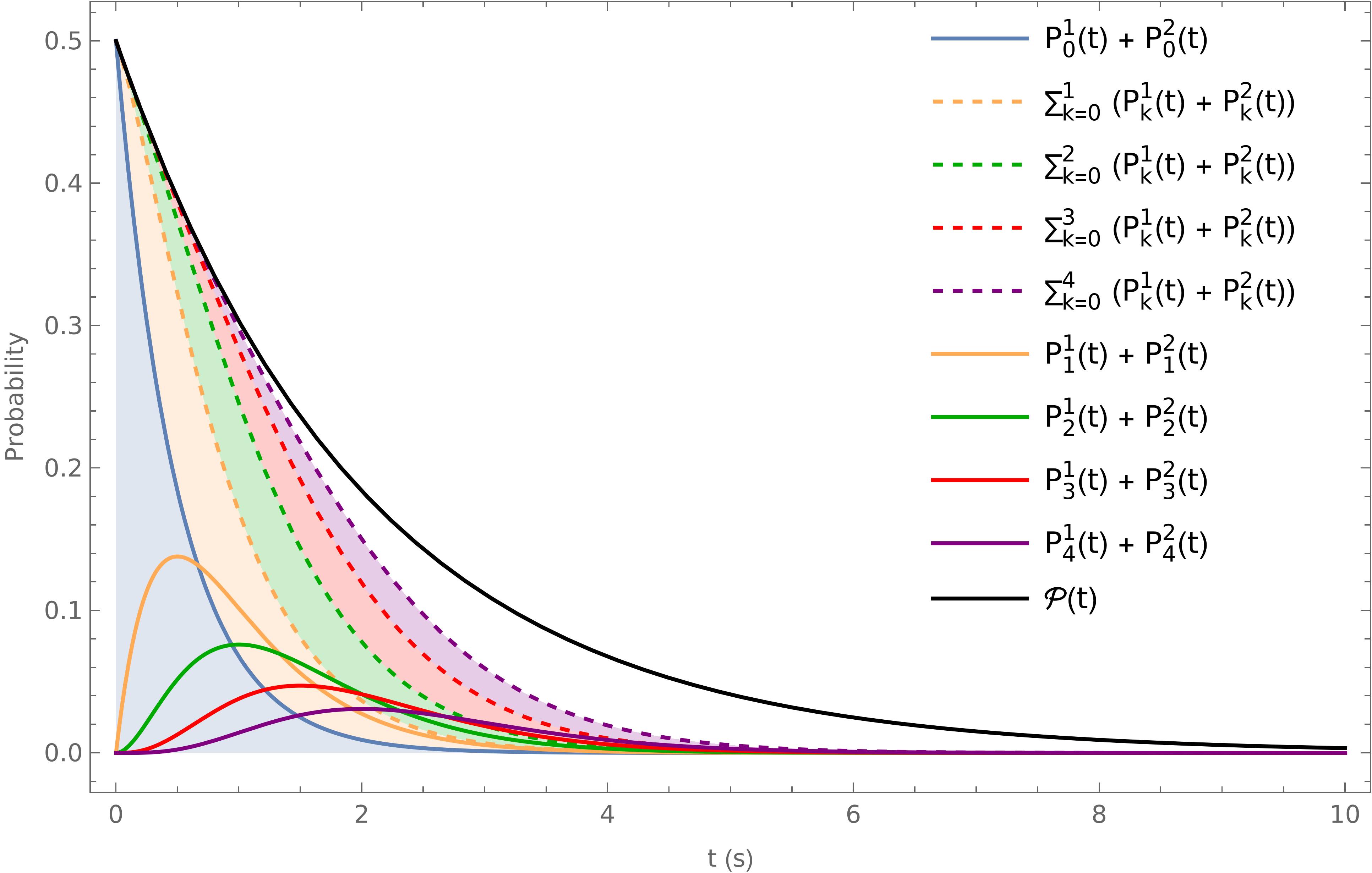}
    \caption{Contributions, $P_k^l(t)= \frac{e^{-t \Gamma} t^{k}\Gamma^{k+1}}{k!}(\frac{R_l}{\Gamma})\left(\frac{\Gamma - R_1 -R_2}{\Gamma}\right)^{k}$, to the total scattering probability, $\mathscr{P}(t)$. $P_k^l$ indicates the $k$th term in the probability distribution sum (for scattering events preceded by $k$ self-scattering events) for rate $R_l$. The probability contributions are taken from two constant scattering rates, $R_1 = 0.2 \ \mbox{s}^{-1}$ and $R_2 = 0.3 \ \mbox{s}^{-1}$. In the probability expressions, $\Gamma = 2$. The terms are plotted as solid lines, while their contribution to the total sum is shown as dashed lines bounding a shaded area, e.g., the orange dashed curve indicates the sum of the solid blue and solid orange curves. $\mathscr{P}(t)$ is shown as a solid black line. 
    }
    \label{fig:ConstRateProbSum}
\end{figure}

The total distribution of free-flight times ending in type 1 scattering will be a collection of variables chosen from these distributions weighted by their respective scattering probability, i.e., a \emph{mixture distribution} of all possible Erlang distributions. The PDF of a mixture distribution is a sum of the component PDFs, weighted by their probabilities: 
\begin{equation}\label{eqn:erlang_series}
    \begin{array}{cc}
      P^1(t) = \frac{R_1}{\Gamma}(e^{-t\Gamma} \Gamma)+(\frac{\Gamma -R_1 -R_2}{\Gamma})(\frac{R_1}{\Gamma})(e^{-t\Gamma}t \Gamma^2) \\[4pt]
      +(\frac{\Gamma -R_1 -R_2}{\Gamma})^2(\frac{R_1}{\Gamma})(\frac{1}{2}e^{-t\Gamma}t^2 \Gamma^3) + ...\\[4pt]
    = \sum\limits_{k=0}^\infty \frac{e^{-t \Gamma} t^{k}\Gamma^{k+1}}{k!}(\frac{R_1}{\Gamma})\left(\frac{\Gamma - R_1 -R_2}{\Gamma}\right)^{k}.
    \end{array}
\end{equation}
Fig.~\ref{fig:ConstRateProbSum} shows the contributions to the total scattering probability of the terms in this sum, $P_k^l(t)$,  corresponding to collisions preceded by different numbers of self-scattering events, $k$, for a mixture distribution with two rates, $R_1 = 0.2 \ \mbox{s}^{-1}$ and $R_2 = 0.3 \ \mbox{s}^{-1}$, with $l=1$ and $l=2$.

The series $P^1(t)$ may be identified as an exponential series, and summed analytically: 
\begin{equation}
    \begin{array}{cc}
    \sum\limits_{k=0}^\infty \frac{e^{-t \Gamma} t^{k}\Gamma^{k+1}}{k!}(\frac{R_1}{\Gamma})(\frac{\Gamma - R_1 -R_2}{\Gamma})^{k} \\[4pt]
    = (\frac{R_1}{\Gamma})\Gamma e^{-t \Gamma} e^{\left(\frac{\Gamma -R_1 -R_2}{\Gamma}-1\right)} \\[4pt]
    = R_1 e^{-(R_1 + R_2)t}.
    \end{array}
\end{equation}
Notice that the arbitrary, constant probability rate, $\Gamma$, cancels out, leaving only scattering rates $R_1$ and $R_2$.

We may find a similar series for $R_2$ and add this to the $R_1$ distribution to obtain the total mixture distribution ($P^1 + P^2$): 
%
\begin{equation}
    (R_1 + R_2) e^{- (R_1+R_2)t} = \frac{1}{\tau} e^{\frac{-1}{\tau}t}.
\end{equation}
Thus, for the case with constant scattering probabilities, the final distribution of free-flight times produced by self-scattering will be exponential, and will have the same relaxation time as the one given by Matthiessen's rule for the two scattering types: $R_{total} = R_1 + R_2$.
This may be extended to any number of constant rate scattering mechanisms.

Somewhat surprisingly, a similar argument may be made for time-dependent scattering rates. In the time-dependent case, each term of the constant scattering rate sum in the constant case, Eqn.~\ref{eqn:erlang_series}, is replaced by a sum of Erlang-like terms, whose final expression is given by Eqn.~\ref{eqn:Pkl}. For the sake of readability, the derivation for the time-dependent case is given in \ref{appendixB}.
In that derivation, we find that for $N$ time-dependent scatterings rates, $R_l$, with $l=1,2,3,\ldots,N$, the total probability of scattering at a given time, $t$, is given by:
\begin{align}
   P(t) &= \sum_{l=1}^{N}P^l (t) \\ \nonumber
    &=\sum_{l=1}^{N} R_l(t) \exp{-\int_0^t \sum\limits_{j=1}^N R_j(t_1) \dd t_1}.
\end{align}
Thus, for both the time-constant and time-dependent relaxation time cases, the analytical expression for the scattering probability, $\mathscr{P}(t)$, Eqn.~\ref{AnalyticalExpression}, is recovered.

\section{Discussion}\label{sec:discuss} 
\begin{figure}
    \centering
    \includegraphics[width=\linewidth]{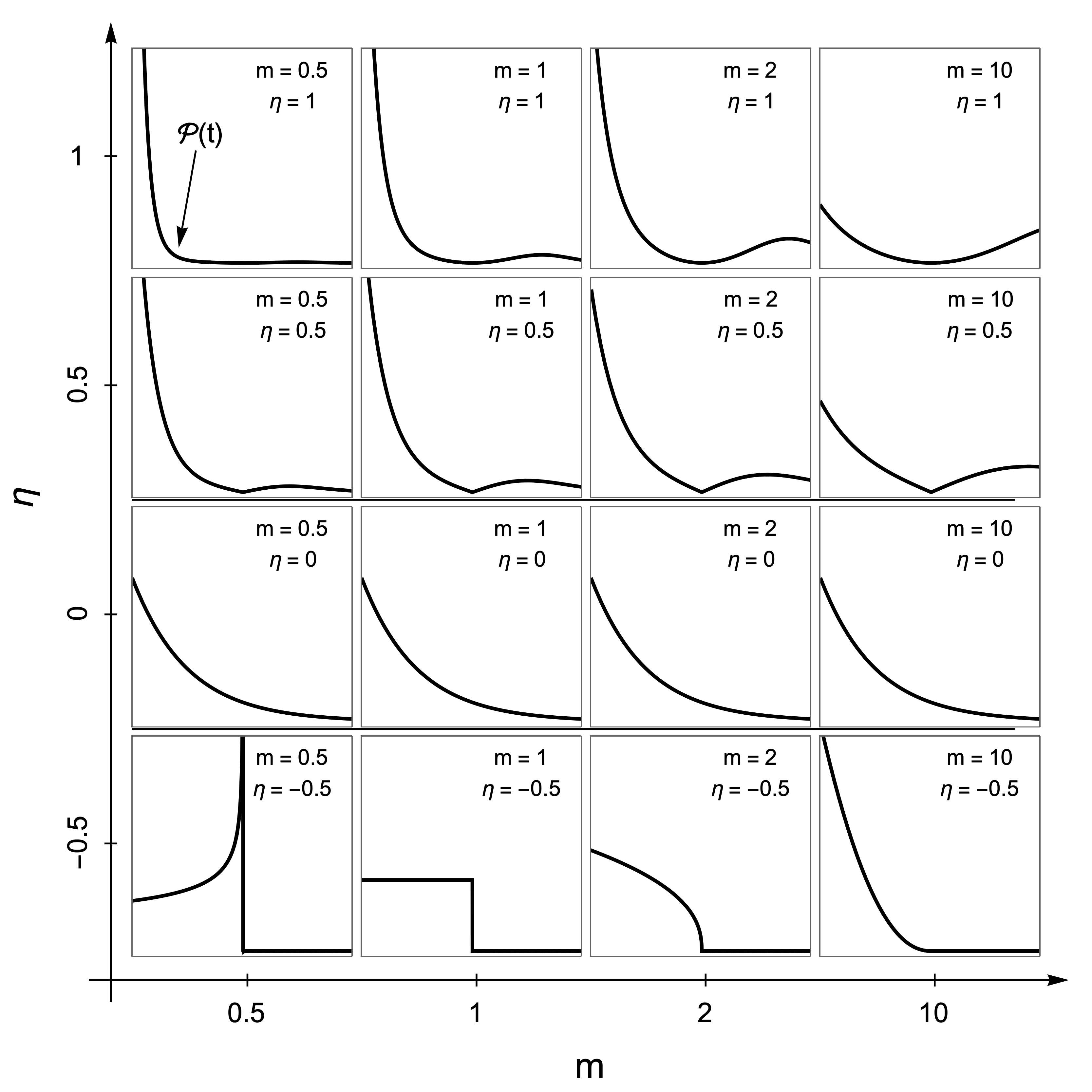}
    \caption{Plots of $\mathscr{P}[\epsilon(t)] $ for a scattering rate, $R[\epsilon(t)] = (\epsilon(t))^\eta$, with nondimensionalized, parabolic dispersion, $\epsilon(t) = (k_0-\alpha t)^2/m$, where $k_0 = 2$ and $\alpha = 1$, for various effective mass values, $m$, and exponents, $\eta$. The constant rate case, $R_0[\epsilon(t)] = 1$, occurs at $\eta = 0$, making $\mathscr{P}[\epsilon(t)]$ the same exponential distribution for any value of $m$. The elastic scattering mechanisms included in our simulations have $\eta = 0.5$. Note that there are many possible forms for $\mathscr{P}[\epsilon(t)]$, even within this simple parabolic dispersion and exponential rate function model.}
    \label{fig:scP_grid}
\end{figure}

The scattering probability expression $\mathscr{P}[\boldsymbol{k}(t)]$ is of fundamental importance in semi-classical Monte-Carlo transport code. It encodes both the scattering modes (through $R[\boldsymbol{k}(t)]$), and the dispersion of the charged particle (through the quasi-particle trajectory given by $\boldsymbol{k}(t)$). It thus captures a lot of the physics of scattering, allowing it to functionally represent the scattering term from the BTE in Monte Carlo Charge Transport simulations. 
Fig.~\ref{fig:scP_grid} gives an example of how even slightly varying parameters in a simple dispersion and rate model can drastically change the shape of $\mathscr{P}[\boldsymbol{k}(t)]$. In this figure, the energy dispersion is chosen to be quadratic ($\epsilon (t) = \frac{(k_0 - \alpha t)^2}{m}$) where $m$ is the effective mass of the dispersion. We make the dispersion explicitly time dependent through the time dependence of the wavevector's trajectory through the Brillouin Zone. The scattering rate is given by $R \propto \epsilon^\eta$, so $\eta$ controls the scattering dependence. We see that when $\eta =0$, $\mathscr{P}$ is exactly exponential. When $\eta$ is larger, we see a significant deviation from an exponential probability for scattering. Likewise when we increase the effective mass, the scattering probability takes on various forms.

The dependence of $\mathscr{P}[\boldsymbol{k}(t)]$, Eqn.~\ref{AnalyticalExpression}, on the full dispersion and path dependent scattering rates makes it computationally expensive to include in its full form in Monte Carlo simulations. 
For this reason, approximations for either the dispersion or the scattering rates or both are often made. In particular, the approximation of an exponential scattering probability in place of the full $\mathscr{P}[\boldsymbol{k}(t)]$ expression is made when constant relaxation times are used to represent the scattering mechanisms. 
The self-scattering technique can account for the energy- and time-dependence of various scattering rates and avoid the over-simplification of the constant relaxation time approximation (CRTA). However, different forms of scattering may be more or less successful in capturing the analytical scattering probability under the self-scattering procedure, as we will explain.

Some features of the probability distribution function created by self-scattering in a Monte Carlo simulation warrant further discussion and lead to a potential limitation of the self-scattering technique. In the semi-classical Monte Carlo method, the probability distribution function, $\mathscr{P}[\boldsymbol{k}(t)]$, Eqn. \ref{AnalyticalExpression}, is re-evaluated at each time-step or free-flight of the simulation. At first glance, this makes each scattering step seem like a memoryless process in time. However, the probability function is implicitly dependent on time through the energy and momentum dispersion of the quasi-particle. The scattering probability is affected by the total time elapsed as the quasi-particle explores more of momentum space over the course of the simulated trajectory. The quasi-particle has a higher probability of reaching states farther away from the starting point after longer times in the simulation. 

\begin{figure}
    \centering
    \includegraphics[width=\linewidth]{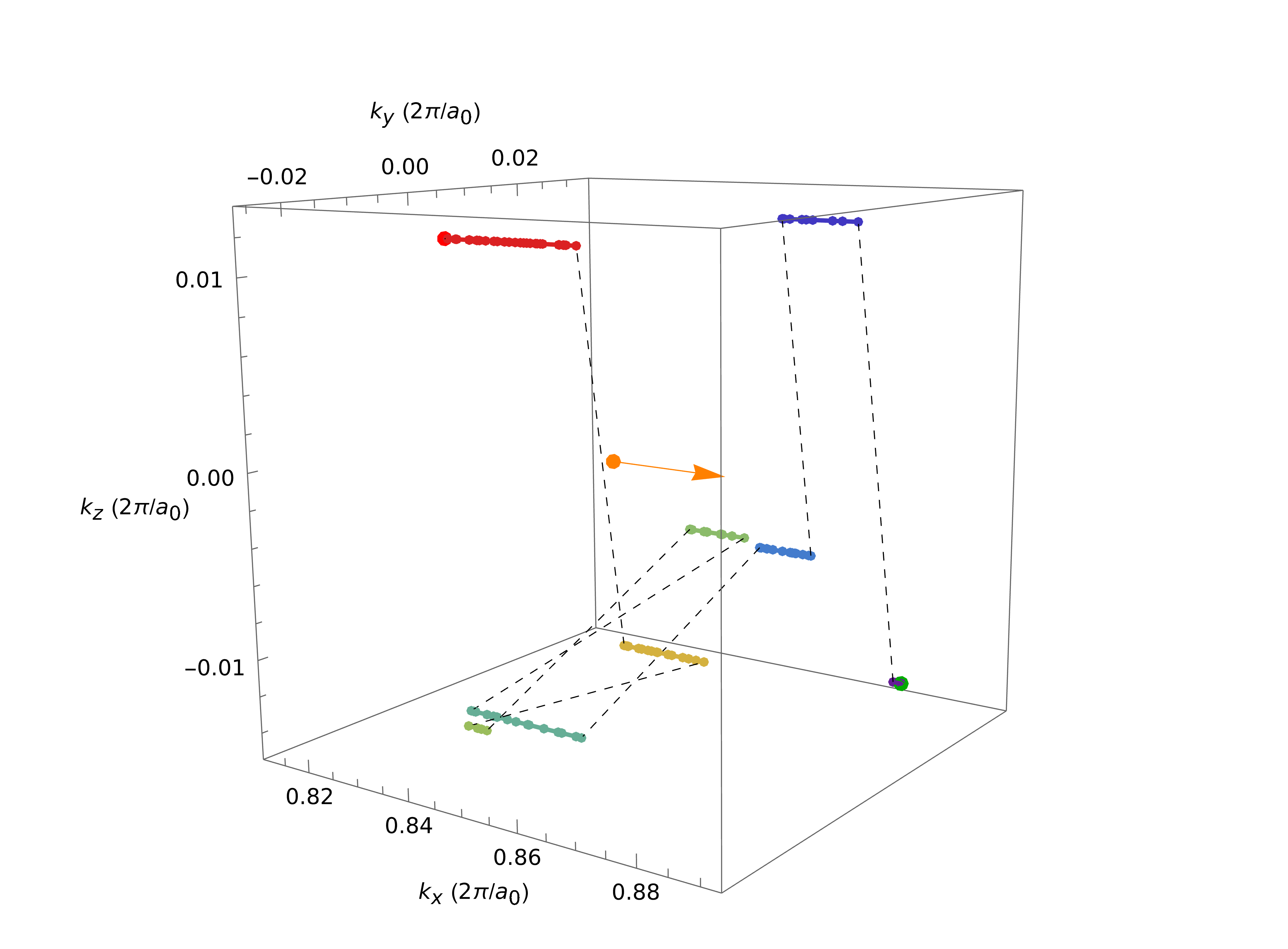}
    \caption{Illustration of a carrier's simulated path through $k$-space in units of $2\pi/a_0$. In the model, $a_0 = 5.43095$ \AA. The corresponding analytical probability expression is shown in Fig.~\ref{fig:kprob}. Elastic acoustic phonon and ionized impurity scattering were simulated using the expressions given in \ref{appendixA}. Dots represent collision events or self-scattering events. The colored lines represent free-flights, while the dashed lines represent scattering transitions. The larger orange dot designates the center of the band minimum, and the orange arrow indicates the direction of the applied electric field. The carrier travels in the direction opposite the applied field during free flight. The initial and final positions are points highlighted in green and red respectively.} 
    \label{fig:kpath}
\end{figure}

\begin{figure}
    \centering
    \includegraphics[width=\linewidth]{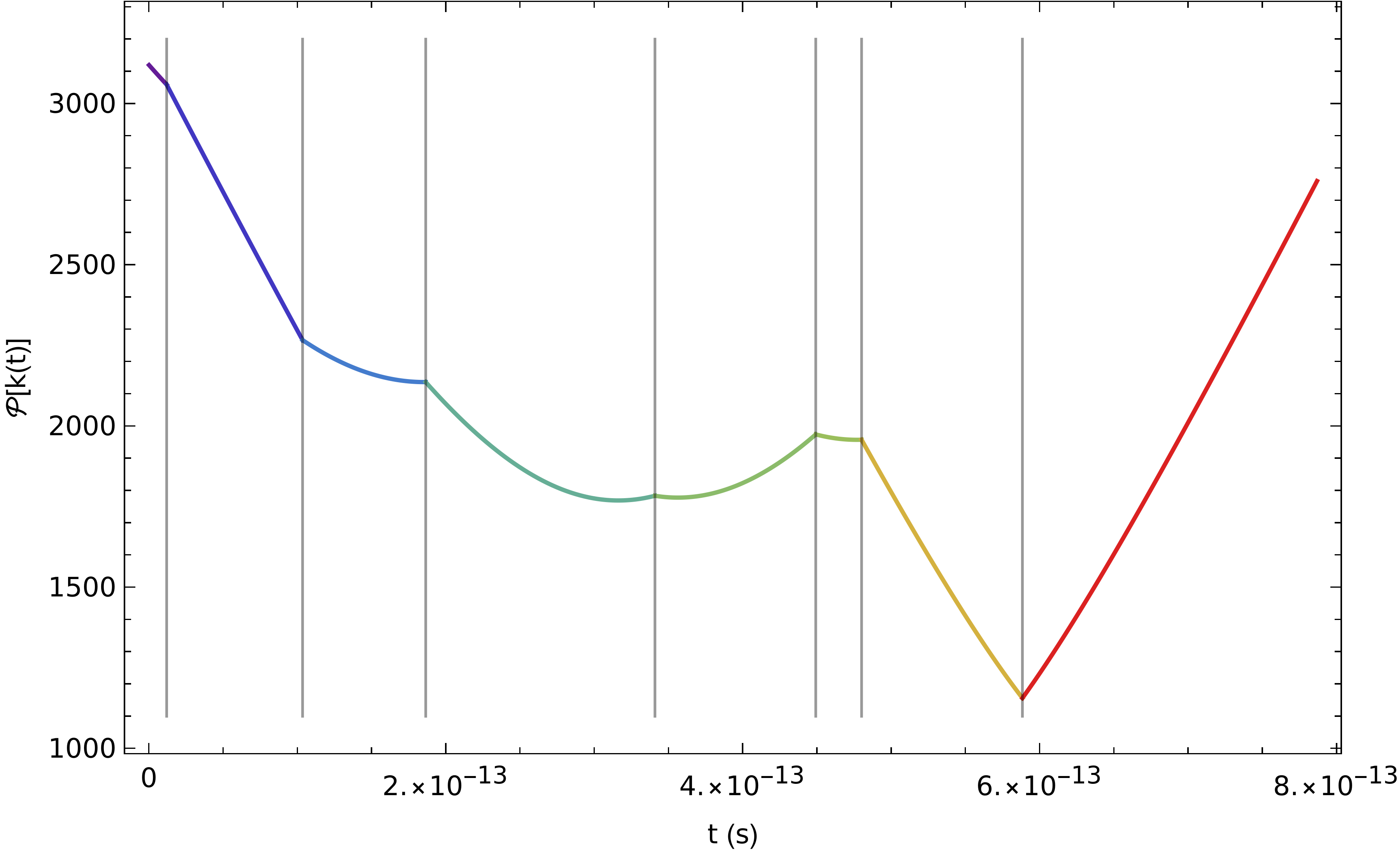}
    \caption{The probability distribution function, $\mathscr{P}[\epsilon[\boldsymbol{k}(t)]]$, corresponding to the simulated path in Fig.~\ref{fig:kpath}. Free-flights between real scattering events are shown in different colors matching the colors of the free flights in Fig.\ref{fig:kpath}. Note that different areas of the probability curve, which corresponds to an $\eta = 0.5$ curve in Fig.~\ref{fig:scP_grid}, are sampled as the quasi-particle travels through k-space. $\mathscr{P}[\epsilon[\boldsymbol{k}(t)]]$ is approximately continuous in this case because of energy conservation.}
    \label{fig:kprob}
\end{figure}

After the simulated trajectory, if we piece together the probability as a function of total elapsed time, we obtain a piecewise function made up of a series of $\mathscr{P}[\boldsymbol{k}(t)]$ functions, having different shapes depending on where the free-flight began in $k$-space. This piecewise scattering probability function will be unique to a given trajectory in $k$-space, because it depends on the electron's path. This is illustrated in Figs.~\ref{fig:kpath} and \ref{fig:kprob}, which respectively show the carrier's path through k-space for a simulation with elastic scattering mechanisms with an energy dependence of $\epsilon^{1/2}$ and the corresponding scattering probability, $\mathscr{P}[\epsilon[\boldsymbol{k}(t)]]$ as a function of time. Corresponding areas of free-flights are shown in the same color. The carrier trajectory in Fig.~\ref{fig:kpath} begins at the large green point on the right side of the figure and ends at the large red point on the left. The direction of the applied field is shown as an orange arrow from the center of the band minimum at the orange point in the figure. Note that the scattering probability in Fig.~\ref{fig:kprob} is approximately continuous because of energy conservation in elastic scattering.
The dependence of the scattering probability, $\mathscr{P}$, on the total time is also reflected in the treatment of the time variable in our derivation of the time-dependent expression, e.g., Eqn.~\ref{eqn:P1l}. 

The dependence of $\mathscr{P}[\boldsymbol{k}(t)]$ on the total simulation time through the quasi-particle's path points to cases of energy dispersions or scattering rates where the self-scattering technique might fail. If the simulation uses scattering rates which are highly peaked in $k$-space and time near the beginning of the trajectory but very small everywhere else, longer free-flight times could be selected by self-scattering which would entirely miss these features and allow the carrier to continue traveling under the influence of the applied field without scattering. Such trajectories would then not be an accurate sampling of the system. 

\begin{figure}
    \centering
    \includegraphics[width=\linewidth]{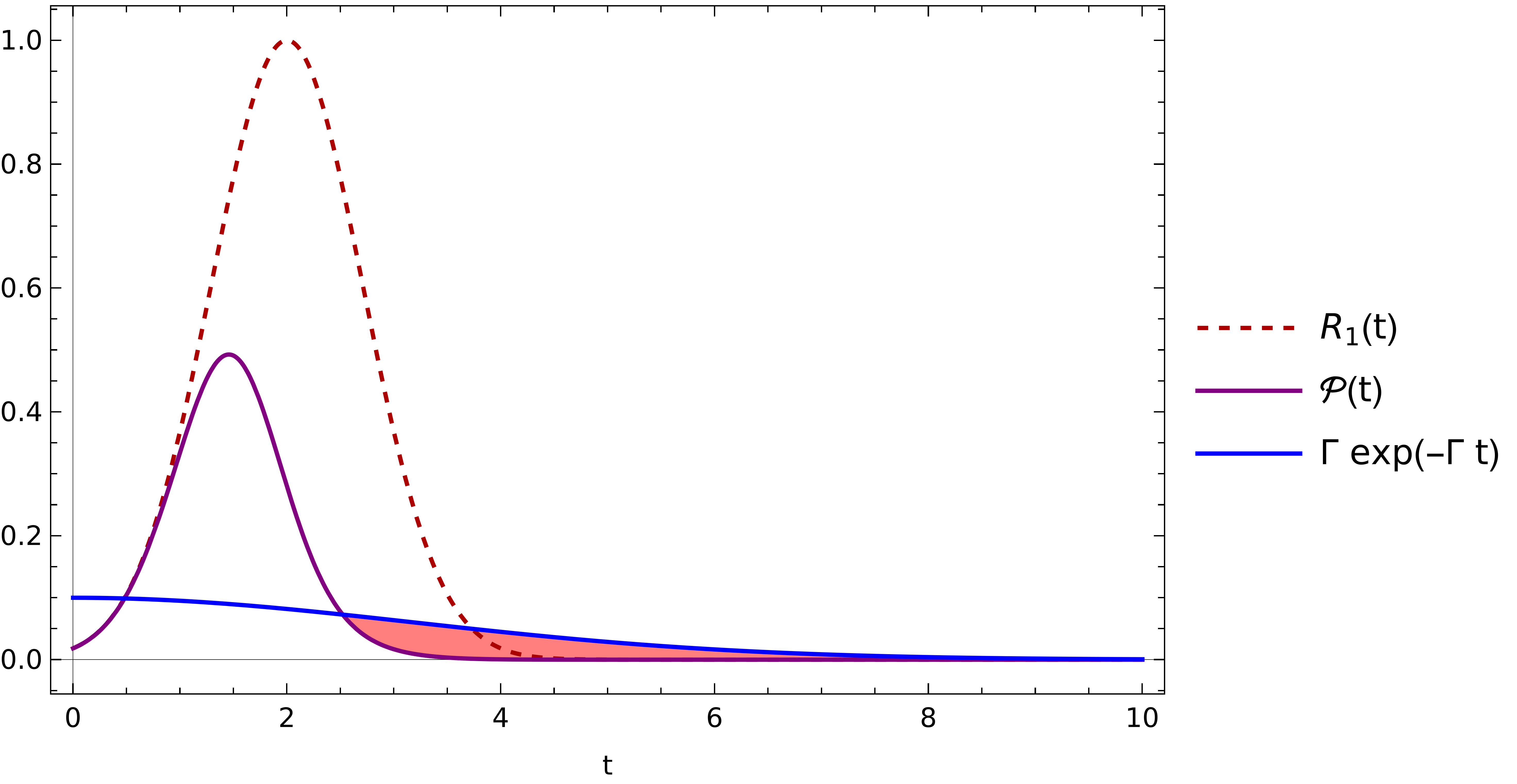}
    \caption{Plots of the total scattering probability, $\mathscr{P}(t)$, shown in purple, for one scattering rate, $R_1(t)=\exp(-(t-2)^2)$, shown as a red dashed curve, and an exponential distribution with $\Gamma = 0.1$, shown in blue. In this example, $\mathscr{P}(t) = \operatorname{exp}\{-(-2 + t)^2 + 
 (1/2 )(-(1/e^4) + \operatorname{exp}[-(-2 + t)^2 - 2 \sqrt{\pi} \operatorname{erf}(2) + 
    2 \sqrt{\pi} \operatorname{erf}(2-t)]\}$.
    Times corresponding to the overlapping area shaded in pink are still likely to be selected as free-flight times in self-scattering based on this exponential distribution, but it is unlikely for scattering to occur at these times, because the highest values of $\mathscr{P}(t)$ occur at earlier times. Selection of appropriate values for $\Gamma$ is important to avoid cases such as this.}
    \label{fig:GaussianandSmallGamma}
\end{figure}

For example, consider a scattering rate which is a Gaussian distribution in time. For this rate, the scattering probability also appears Gaussian-like.  
If the exponential parameter used in self-scattering, $\Gamma$, is taken to be too small, cases where the free-flight time selected will miss the peak may occur relatively frequently. This is illustrated in Fig.~\ref{fig:GaussianandSmallGamma}. In this figure, we have a Gaussian scattering rate plotted in time (in red), which is peaked at $t=2$, the corresponding $\mathscr{P}(t)$ function, which is peaked at $t=1.45727$, plotted in purple, and an exponential distribution with an disproportionately small value of $\Gamma = 0.1$, plotted in blue. 
Here we can see that there is a significantly large relative probability of choosing free-flight times that completely miss the bulk of the scattering process, illustrated by the shaded area of the plot. 

Even if the scattering rate increases in time as the quasi-particle travels away from the starting position, taking too small of an exponential parameter could lead to outliers in the data where the electron traveled an unrealistic distance in $k$-space because of the large free-flight time selected. This could cause $k$-space to be under-sampled and an unrealistic spread in momentum and energy in the simulation.
A similar issue is brought up by Jacoboni et al. in their discussion of the rejection technique for selection of random variables from a distribution, in Appendix A.2.b of reference \cite{Jacoboni1983}.

While selecting a larger value for the exponential parameter $\Gamma$ may help avoid some of these issues, selecting too large of a value for $\Gamma$ may lead to similar problems. If $\Gamma$ is large enough that self-scattering will be selected instead of a real scattering event across realistic regions of k-space for the carrier to travel, there will be not only a lack of statistics for the real scattering events, leading to computational inefficiency, but the small free-flights chosen by self-scattering may add up to larger unrealistic total free-flight times when the self-scattering events are removed from the simulation in post-processing. It's possible in this way to re-create the same issue of unrealistically long free-flight times which may occur in choosing too small of a value for $\Gamma$. Care must be taken to chose a value for $\Gamma$ which is proportionate to the scattering rates included in the simulation.

Thus, we see that while self-scattering in theory will capture the time dependence of various scattering mechanisms, as shown in \ref{appendixB}, it has the potential to distort the full time and energy dependence of $\mathscr{P}[\boldsymbol{k}(t)]$ through its implementation in the code if the scattering probability has unusual features from the scattering rate or dispersion expressions, or if the value for the exponential parameter is not selected carefully.
In practice, it appears that most ills can be accounted for (although by accruing serious computational cost) by increasing $\Gamma$ sufficiently that there are enough self-scattering events to follow $\mathscr{P}[\boldsymbol{k}(t)]$ sufficiently well, while still keeping $\Gamma$ in the order of the scattering rates, although more work is required to determine whether this is always the case. Increasing the number of trajectories simulated also has the potential to improve the accuracy of the free-flight time distribution sampling within self-scattering. 
Jacoboni et al. \cite{Jacoboni1983} recommend setting the value of the exponential parameter, $\Gamma$ to a value which will be greater than the sum of the scattering rates in any area in $k$-space the quasi-particle is expected to travel in a simulation \cite{Jacoboni1983}. For higher energy simulations, they divide momentum space up into smaller areas where different exponential variables are used in order to increase simulation efficiency while ensuring the parameter is great enough to hopefully avoid error. The reader is referred to their discussion of the self-scattering technique in \cite{Jacoboni1983} for more details about this method for selection of the exponential parameter.

More work is required to determine the optimal exponential parameter and ensemble size selection for various shapes of scattering rates and dispersion functions. Additionally, the bounds of the error that the self-scattering technique could introduce into charge transport estimators produced by the simulation (such as drift velocity and diffusion coefficients) has yet to be explored. Many other aspects of the process and potential difficulties likely could appear with further exploration, since this technique has not previously been well characterized.

\section{Conclusion}
We have shown that the self-scattering technique for semi-classical Monte Carlo charge transport simulations recreates the scattering rates and probability functions intended for the simulation. 
The simulated total scattering rate and ratios of scattering events were found to be consistent with the assumed rates within simulations implementing self-scattering in Section~\ref{sec:expverf}. In Section~\ref{sec:dists} and \ref{appendixB}, the probability distribution created by self-scattering was analytically shown to be the same as the probability distribution function dependent on the scattering rates, both for rates constant in time and rates which were dependent on time. We found that self-scattering produces a mixture distribution which is a series of weighted Erlang or Erlang-like functions, in which each term represents the collection of scattering events proceeded by a given number of self-scattering events, and that this series may be summed to produce the scattering probability expression $\mathscr{P}$, given by Eqn.~\ref{AnalyticalExpression}. 

While these explorations and analyses indicate that self-scattering is a viable technique for use in Monte Carlo simulations of charge transport, in Section~\ref{sec:discuss}, we discussed some potential limitations of the technique. In particular, because of the complicated energy, momentum, and time-dependence of the scattering probability expression, $\mathscr{P}$, selecting too large or too small of a value for the exponential parameter, $\Gamma$, used in the exponential distribution to select free-flight times in self-scattering, can lead to inefficient or unrealistic simulations.

Clarity about such processes for implementation of scattering models is important to ensure accurate simulations and to improve on simulation models. The self-scattering technique has been shown to be a valid technique and useful for its ease of implementation and efficiency in Monte Carlo carrier transport simulations. However, because of the complicated intricacy of energy- and time- dependent scattering mechanisms, there is still work to be done towards understanding the full impact of implementing self-scattering. 

\section*{Acknowledgements}
The authors would like to thank Roman Mecholsky for providing valuable discussion concerning the proofs in Section 5 and Appendix B as well as many discussions of aspects of those derivations. Additionally the authors would like to thank VSL for financially supporting this research.


\appendix

\section{Scattering Rates}\label{appendixA}
A spherical-parabolic dispersion expression was used to determine these scattering rates. For more details about these expressions, please refer to Jacoboni et al. \cite{Jacoboni1983, Jacoboni1989}.
\begin{enumerate}
    \item[1)]{\emph{Electron-Acoustic Phonon}}
    
An elastic acoustic phonon scattering rate as a function of energy may be given by: 
\begin{equation}\label{eqn:El_Ac}
    R_{ac}^{JR}(\epsilon) = \frac{\sqrt{2} m^{3/2} K_B T \mathscr{E}_1^2}{\pi \hbar^4 u_l^2 \rho} \epsilon^{1/2}.
\end{equation}

Inelastic acoustic phonon scattering rates as a function of energy may be given by: 
\begin{equation}\label{eqn:Inel_Ac_abs}
   R_{ac,abs}^{JR}(\epsilon) = \frac{m^{1/2}(K_B T_0)^3\mathscr{E}_1^2}{2^{5/2}\pi \rho u_l^4\hbar^4}\epsilon^{-1/2}(F_1(x_{2a})-F_1(x_{1a})),
\end{equation}
for absorption, and
\begin{equation}\label{eqn:Inel_Ac_em}
    R_{ac,em}^{JR}(\epsilon) = \frac{m^{1/2}(K_B T_0)^3\mathscr{E}_l1^2}{2^{5/2}\pi \rho u_l^4\hbar^4}\epsilon^{-1/2}(G_1(x_{2e}) - G_1(x_{1e})),
\end{equation}
for emission processes.
In these expressions, $m$ is an average effective mass for the band model, $K_B$ is the Boltzmann constant, $T_0$ is the system temperature, $\rho$ is the material density, $u_l$ is the longitudinal speed of sound in the material, $x$ is a dimensionless parameter dependent on the phonon wavevector $q$, and $F(x)$ and $G(x)$ are functions dependent on the number of phonons in the material given by the Bose-Einstein distribution $N_q(x)$:
\begin{equation}
    F_1(x) = \int_0^x N_q(x')x'^{2} dx',
\end{equation}

\begin{equation}
     G_1(x) = \int_0^x [N_q(x') + 1] x'^{2} dx',
\end{equation}
where:
\begin{equation}
    N_q(x) = \frac{1}{e^x-1}.
\end{equation}

In section~\ref{sec:expverf}, the energy-dependent simulation utilizes the inelastic rates, Eqns.~\ref{eqn:Inel_Ac_abs} and \ref{eqn:Inel_Ac_em}, while the energy-independent simulation uses the elastic phonon rate, Eqn.~\ref{eqn:El_Ac}, with $\epsilon =1$.
\item[2)]{\emph{Electron-Ionized Impurity}:}
For ionized impurity scattering, the Conwell and Weisskopf model \cite{Conwell1950} as implemented in \cite{Jacoboni1983} was used:
\begin{equation}\label{eqn:II}
    R_{ion}^{CW}(\epsilon) = \pi n_I Z^2 b^2 (2/m)^{1/2} \epsilon^{1/2}
\end{equation}
In this expression, $n_I$ is the number of impurities per unit volume, $Z$ is the number of impurity charge units and $b = (\frac{3}{4 \pi n_I})^{1/3}$ is the mean distance between impurities. The Conwell and Weisskopf approach uses a scattering potential $\mathscr V(r) = \frac{Z |e|^2}{\kappa r}$ cut off after the mean distance between impurities, or impact parameter, $b$. The parameter $\kappa$ is the dielectric constant of the material and $|e|$ is the magnitude of the electron charge.

In section~\ref{sec:expverf}, the full energy-dependent Eqn.~\ref{eqn:II} was used for the energy-dependent simulation, while in the energy-independent simulation, Eqn.~\ref{eqn:II} was used with $\epsilon = 1$.
\end{enumerate}

\section{Time-Dependent Probability Distribution}\label{appendixB}

The more general, time-dependent distribution may be approached by partitioning the data by the number of self-scattering events in a given free-flight and summing a series for each scattering type. Each of these series corresponds to a single term in the scattering type 1 series we created for the constant case. Dividing the sum in this way gives insight into the contributions of each $n$-self-scattering event distribution to the final free-flight time probability density function.

We will approach the time-continuous distribution by discretizing time into small, equal units indexed by: $i = 1,2,3,...,\infty$. We will pick free-flight times from the set of all intervals chosen from an underlying distribution with probability $p_i$ of choosing a time interval indexed by $i$.

Suppose we have $N$ scattering mechanisms. Define the probability of choosing the $l$th scattering mechanism at a time interval $n$ as $q_{nl}$. With this notation, the probability of self-scattering occurring at interval $n$ is given by $\sum\limits_{j=1}^N (1-q_{nj})$. Let $P_k^l(n)$ be the probability of scattering with mechanism $l$ at time interval $n$, preceded by $k$ self-scattering events. The total probability of scattering with mechanism $l$ at time step $n$ will be the sum of all of the possible self-scattering combinations: $P^l(n) = \sum\limits_{k=0}^\infty P_k^l(n)$. Let us write out this sum. 
In self-scattering, the time interval at which scattering occurs is chosen, then the scattering mechanism is retroactively selected based on its probability evaluated at that particular time. Assuming these processes are independent, the \emph{first term} ($k=0$) of the series that describes all events that will eventually scatter at the $n$th interval with the $l$th mechanism will be the product of the probability that the time interval $n$ is chosen, times the probability that scattering mechanism $l$ is chosen at that time, $P_0^l(n) = p_n q_{nl}$.

The second term, with 1 self-scattering event, may be written as: 
\begin{equation}\label{eqn:P1l}
   P_1^l(n) = \sum\limits_{i=1}^{n-1} p_i p_{n-i}(1-\sum\limits_{j=1}^N q_{ij}) q_{nl}. 
\end{equation} 
Let us unpack this expression. It represents two collision events, a self-scattering event occurring at time $i$ (between 0 and $n$), and a true scattering event occurring at a time $n$. There are four separate terms which combine to create the total probability: the probability of selecting free-flight time $i$ before the self-scattering event, $p_i$; the probability of selecting the remaining free-flight time before the true scattering event at $n$, $p_{n-i}$; the probability that self-scattering occurs at the first time interval selected, $(1-\sum\limits_{j=1}^N q_{ij})$; and the probability that scattering type $l$ occurs at the total time $n$, $q_{nl}$. 

The time dependence of the scattering mechanisms comes from the variation of the scattering probability through $k$-space in time, and so will depend on the total elapsed time (here represented by $n$). The self-scattering probability term, $(1-\sum\limits_{j=1}^N q_{ij})$, is the probability that no other type of scattering occurs at time interval $i$. The sum over $i$ includes the terms $i=1$ through $i = n-1$ to take into account all of the available time intervals before $n$ where self-scattering could occur.

Following the same scheme, the third term in the series, $P_2^l(n)$, may be written as:
\begin{equation}
    P_2^l(n) = \sum\limits_{i_1=1}^{n-2} \mspace{9mu} \sum\limits_{i_2=i_1+1}^{n-1} p_{i_1} p_{{i_2}-{i_1}} p_{n-{i_2}} (1-\sum\limits_{j=1}^N q_{{i_1}j}) (1-\sum\limits_{j = 1}^N q_{{i_2}j}) q_{nl}.
\end{equation}
Similarly the fourth:
\begin{align}
    P_3^l(n) = & \sum\limits_{{i_1}=1}^{n-3} \mspace{9mu} \sum\limits_{{i_2}={i_1}+1}^{n-2} \mspace{9mu} \sum\limits_{{i_3}={i_2}+1}^{n-1} p_{i_1} p_{{i_2}-{i_1}} p_{{i_3}-{i_2}} p_{n-{i_3}} \nonumber \\
    & \times (1-\sum\limits_{j=1}^N q_{{i_1}j}) (1-\sum\limits_{j = 1}^N q_{{i_2}j}) (1-\sum\limits_{j = 1}^N q_{{i_3}j}) q_{nl},
\end{align}
and so on. 

Let us now extend this sum to scattering rates and probabilities which are continuous functions in time. The continuous expression for $P_k^l(T)$ will be a limit of the discrete sum as $n \xrightarrow[]{} \infty$. Let the scattering event occur at a time $T$, let $n$ be the total number of time intervals in time $T$ and $\Delta t$ be the length of a time interval, so that $\Delta t = T/n$. Assuming that the underlying free-flight time distribution is the exponential distribution with parameter $\Gamma$, we will have:
\begin{equation}
    p_i = \Gamma e^{-\Gamma (i \Delta t)} \Delta t,
\end{equation}
for the probability of selecting a time interval $i$. If the scattering rate is $R_j(t)$ for mechanism $j$ then the scattering probability at a time step $i$ will be:
\begin{equation}
    q_{ij} = \frac{R_j(i \Delta t)}{\Gamma},
\end{equation}
and at time step $n$:
\begin{equation}
    q_{n j} = \frac{R_j(T)}{\Gamma}.
\end{equation}
Writing a general expression for the terms in the sum, we will have for $k$ self-scattering events and a final scattering event of type $l$:
\begin{align}\label{eqn:genTermExp}
    P_k^l(T) =
    & \lim\limits_{n \xrightarrow[]{}\infty} \sum\limits_{i_1 = 1}^{n-k}\quad
    \sum\limits_{i_2 = i_1+1}^{n-(k-1)} \ldots \sum\limits_{\substack{i_{k-1} = \\ i_{k-2}+1}}^{n-2}\quad\sum\limits_{\substack{i_k = \\ i_{k-1}+1}}^{n-1} \Gamma^{k+1}e^{-\Gamma (n \Delta t)} \Delta t^{k+1} \nonumber \\
     & \times \quad \frac{R_l(T)}{\Gamma} \prod\limits_{m=1}^k
    \left(1-\sum\limits_{j=1}^N \frac{R_j(i_m \Delta t)}{\Gamma}\right).
\end{align}
Note that the product of the $p_i$ terms becomes a single exponential, because they are each chosen from the same exponential distribution and the indices $i_1 + (i_2 - i_1) + \cdots + (n - i_{k})$ sum to $n$. This expression looks very similar to the constant time case, Eqn.~\ref{eqn:erlang_series}, especially the portion $\Gamma^{k+1}e^{-\Gamma (n \Delta t)} \Delta t^{k+1}$, which comes from the sum of the free-flight time variables.

The difficulty in evaluating \Eqn{eqn:genTermExp} for any given $k$ is the evaluation of the $k$-dimensional product within the $k$ sums. Because the indices of the sums are all different, it complicates the summation. One way to address this is by extending every sum from 1 to $n-1$. We may then compensate for the overcounting by subtracting diagonal terms and dividing by $k!$ to account for duplicates. The summation may then be organized graphically as a $k$-dimensional array of terms, which is easier to evaluate. As an example, the 3rd term in the series ($k=2$) for a case with only one scattering type will be:
\begin{align}
    P_2^1(T) & = R_1 (T) \lim\limits_{n\xrightarrow[]{} \infty} \sum\limits_{i_1 = 1}^{n-2} \mspace{9mu} \sum\limits_{i_2 = i_1+1}^{n-1} \Gamma^2 e^{-\Gamma T} 
    \nonumber \\
    & \quad \times \left(1 - \frac{R_1(i_1 \Delta t)}{\Gamma}\right)\left(1 - \frac{R_1(i_2 \Delta t)}{\Gamma}\right) \Delta t^3 \nonumber \\ 
    & = \Gamma^2 e^{-\Gamma T} R_1(T) \nonumber \\ 
    & \quad  \times\lim\limits_{n\xrightarrow[]{}\infty}
    \left( \frac{1}{2!} \left( \sum\limits_{i_1 = 1}^{n-1} \left( 1 - \frac{R_1(i_1 \Delta t) }{\Gamma}\right)\Delta t \sum\limits_{i_2 = 1}^{n-1} \left( 1 - \frac{R_1 (i_2 \Delta t) }{\Gamma}\right)\Delta t \right. \right. \nonumber \\
    & \quad - \left. \left. \sum\limits_{i = 1}^{n-1} \left(1 - \frac{R_1 (i \Delta t) }{\Gamma}  \right)^2 \Delta t^2 \right) \Delta t \right). 
\end{align}
The diagonal term will go to 0 as $\Delta t \xrightarrow[]{} 0$ because it is of order $O(\Delta t^2)$ in the limit, but the off diagonal terms become integrals. Thus:
\begin{equation}
    P_2^1(T) = \Gamma^2 e^{-\Gamma T} R_1(T) \frac{1}{2!} \left(\int_0^T\left( 1 - \frac{R_1(t)}{\Gamma} \right) \dd t \right)^2,
\end{equation}
and extending this argument:
\begin{equation}\label{eqn:Pkl}
    P_k^l(T) = \frac{\Gamma^{k+1} e^{-\Gamma T}}{k!} \left(\frac{R_l(T)}{\Gamma}
     \right)\left(\int_0^T\frac{\Gamma - \sum_{j=1}^N R_j(t)}{\Gamma} \dd t \right)^k.
\end{equation}
Notice that if the scattering rates are taken to be constant in time in this expression, these terms simplify to the corresponding Erlang distribution terms in the sum, Eqn.~\ref{eqn:erlang_series}, from the constant rate derivation. 

\begin{figure}
    \centering
    \includegraphics[width=\linewidth]{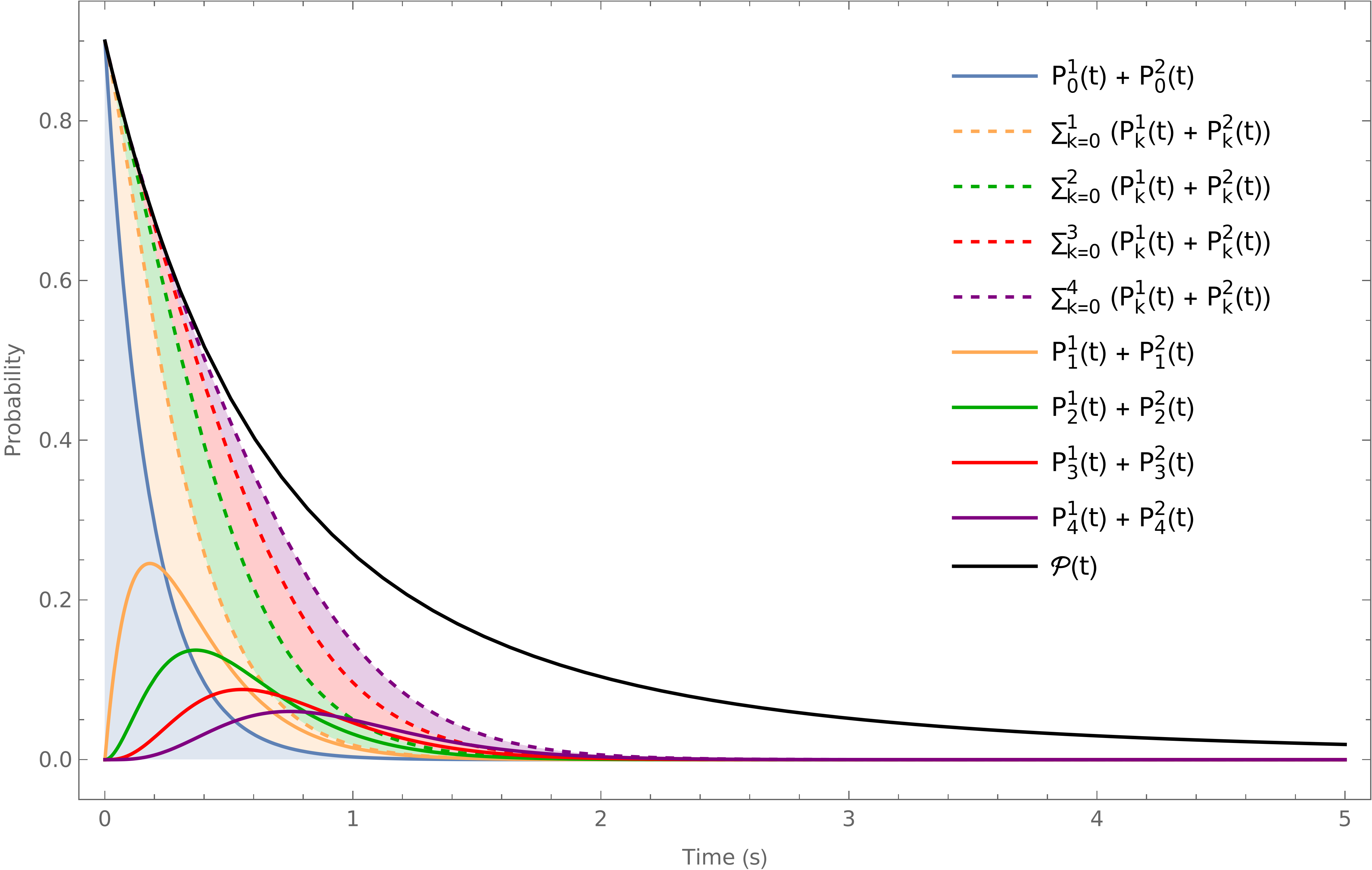}
    \caption{Contribution to the total probability of scattering by terms corresponding to collisions preceded by 0, 1, 2, 3, and 4 self-scattering events shown in Blue, Yellow, Green, Red, and Purple, respectively, as a function of time. Rates 1 and 2 are given by $R_1(t) = 0.2 \exp(-0.2 t)$ and $R_2(t) = 0.7 \exp(-0.7 t)$. $P_k^l(t)$ is given by Eqn.~\ref{eqn:Pkl}. $\Gamma$ in these expressions is set to 5. The dashed lines and shaded areas show the terms added to the preceding terms, while the solid lines show bare functions. The black solid line is the total analytical scattering probability.  As an example, the blue curve indicates the probability for scattering directly without self-scattering, while the dashed yellow curve is the sum of the solid blue curve with the solid yellow curve, indicating the probability of scattering with 0 or 1 self-scatterings.}
    \label{fig:sumplots}
\end{figure}

These terms and their contribution to the final expression are plotted in Fig.~\ref{fig:sumplots} for the case of two exponential scattering rates ($R_1(t) = 0.2 \exp(-0.2 t)$ and $R_2(t) = 0.7 \exp(-0.7 t)$). The total scattering probability is shown as a black curve, while the terms are plotted as solid colored curves. The total sums up through the contribution of the corresponding terms are plotted as dashed colored curves with the additional contribution between terms shaded in. Note that the sum very quickly converges to the total scattering probability at early times in this example, but takes more terms to converge at later times. While the total probability appears exponential-like in this case, the shape of the function could be wildly different, as indicated in Fig.~\ref{fig:scP_grid}.

Summing the partial terms we obtain the probability expression for one scattering mechanism: 
\begin{align}
    P^l (T) & = P_0^l(T) + P_1^l(T) + P_2^l(T) + \ldots \nonumber \\
    & = e^{-\Gamma T} R_l (T) + \Gamma e^{-\Gamma T} R_l(T) \left( \int_0^T \left( 1 - \sum\limits_{j = 1}^N \frac{R_j(t)}{\Gamma} \dd t \right) \right) \nonumber \\
    & \quad + \Gamma^2 e^{-\Gamma T} R_l(T) \frac{1}{2!} 
    \left( \int_0^T \left( 1 - \sum\limits_{j=1}^N\frac{R_j(t)}{\Gamma}  \right) \dd t \right)^2 + \ldots \nonumber \\
    & = R_l(T) e^{-\Gamma T} e^{\Gamma T} \exp{- \int_0^T \sum\limits_{j = 1}^N R_j(t) \dd t} \nonumber \\
    & = R_l(T) \exp{-\int_0^T \sum\limits_{j=1}^N R_j(t) \dd t}.
\end{align}

Thus, when all the mechanisms are included, we recover the analytic expression for the total scattering probability as a function of time from Matthiessen's rule for both the time independent and dependent scattering rate cases:
\begin{equation}\tag{\ref{AnalyticalExpression}}
    \mathscr{P}(t) = R(t) \exp(-\int\limits_0^t R(t') \dd t').
\end{equation}
%
%

\bibliographystyle{elsarticle-num} 
\bibliography{sources}

\end{document}